
%
%
%
%
\def\bra#1{{\langle#1\vert}}
\def\ket#1{{\vert#1\rangle}}
\def\coeff#1#2{{\scriptstyle{#1\over #2}}}
\def\undertext#1{{$\underline{\hbox{#1}}$}}
\def\hcal#1{{\cal #1}}
\def\sst#1{{\scriptscriptstyle #1}}
\def\eexp#1{{\hbox{e}^{#1}}}
\def\rbra#1{{\langle #1 \vert\!\vert}}
\def\rket#1{{\vert\!\vert #1\rangle}}

\def\nubar{{\bar\nu}}

\def\alr{{A_\sst{LR}}}

\def\evec{{\vec e}}

\def\notcder{{\not\!\! D}}

\def\mn{{m_\sst{N}}}
\def\mns{{m^2_\sst{N}}}

\def\sbar{{\bar s}}

\def\sstw{{\sin^2\theta_\sst{W}}}

\def\pv{{\vec p}}

\def\ppv{{{\vec p}^{\>\prime}}}

\def\qv{{\vec q\,}}

\def\xv{{\vec x}}
\def\xpv{{{\vec x}^{\>\prime}}}

\def\tauv{{\vec\tau}}
\def\sigv{{\vec\sigma}}

\def\sqr#1#2{{\vcenter{\vbox{\hrule height.#2pt
		\hbox{\vrule width.#2pt height#1pt \kern#1pt
			\vrule width.#2pt}
		\hrule height.#2pt}}}}
\def\square{{\mathchoice\sqr74\sqr74\sqr{6.3}3\sqr{3.5}3}}

\def\hcal#1{{\hbox{\cal #1}}}
\def\sst#1{{\scriptscriptstyle #1}}

\def\mpi{{m_\pi}}
\def\mpis{{m^2_\pi}}

\def\mn{{m_\sst{N}}}
\def\mns{{m^2_\sst{N}}}

\def\mro{{m_\rho}}
\def\mros{{m^2_\rho}}

\def\gpnn{{g_{\sst{NN}\pi}}}

\def\Hhat{{\hat H}}

\def\sst#1{{\scriptscriptstyle #1}}
\def\hcal#1{{\hbox{\cal #1}}}
\def\eexp#1{{\hbox{e}^{#1}}}

\def\Hhat{{\hat H}}
\def\That{{\hat T}}

\def\Mhat{{\hat M}}

\def\rohat{{\hat\rho}}

\def\OP{{\hat\hcal{O}}}
\def\mn{{m_\sst{N}}}
\def\mns{{m_\sst{N}^2}}

\def\mpi{{m_\pi}}
\def\mpis{{m^2_\pi}}

\def\qv{{\vec q}}
\def\pv{{\vec p}}
\def\ppv{{{\vec p}^{\>\prime}}}
\def\kv{{\vec k}}

\def\kvs{{kv^{\, 2}}}
\def\xv{{\vec x}}
\def\xpv{{{\vec x}^{\>\prime}}}

\def\rv{{\vec r}}

\def\sigv{{\vec\sigma}}
\def\tauv{{\vec\tau}}

\def\gpnn{{g_{\pi\sst{NN}}}}

\def\rbra#1{{\langle#1\parallel}}
\def\rket#1{{\parallel#1\rangle}}

\def\xivz{{\xi_\sst{V}^{(0)}}}

\def\xivtez{{\xi_\sst{V}^{T=0}}}
\def\xivteo{{\xi_\sst{V}^{T=1}}}

\def\FOS{{F_1^{(s)}}}

\def\GES{{G_\sst{E}^{(s)}}}
\def\GMS{{G_\sst{M}^{(s)}}}

\def\mustr{{\mu_s}}

\def\rhostr{{\rho_s}}

\def\GEn{{G_\sst{E}^n}}
\def\GEp{{G_\sst{E}^p}}
\def\GMn{{G_\sst{M}^n}}
\def\GMp{{G_\sst{M}^p}}

\def\GETEZ{{G_\sst{E}^{\sst{T}=0}}}

\def\GMTEZ{{G_\sst{M}^{\sst{T}=0}}}

\def\lamn{{\lambda_n}}

\def\bra#1{{\langle#1\vert}}
\def\ket#1{{\vert#1\rangle}}
\def\coeff#1#2{{\scriptstyle{#1\over #2}}}
\def\undertext#1{{$\underline{\hbox{#1}}$}}
\def\hcal#1{{\hbox{\cal #1}}}
\def\sst#1{{\scriptscriptstyle #1}}
\def\eexp#1{{\hbox{e}^{#1}}}
\def\rbra#1{{\langle #1 \vert\!\vert}}
\def\rket#1{{\vert\!\vert #1\rangle}}

\def\nubar{{\bar\nu}}

\def\alr{{A_\sst{LR}}}

\def\evec{{\vec e}}

\def\notcder{{\not\!\! D}}

\def\mn{{m_\sst{N}}}
\def\mns{{m^2_\sst{N}}}

\def\sbar{{\bar s}}

\def\sstw{{\sin^2\theta_\sst{W}}}

\def\pv{{\vec p}}

\def\ppv{{{\vec p}^{\>\prime}}}

\def\qv{{\vec q}}

\def\xv{{\vec x}}
\def\xpv{{{\vec x}^{\>\prime}}}

\def\tauv{{\vec\tau}}
\def\sigv{{\vec\sigma}}

\def\sqr#1#2{{\vcenter{\vbox{\hrule height.#2pt
		\hbox{\vrule width.#2pt height#1pt \kern#1pt
			\vrule width.#2pt}
		\hrule height.#2pt}}}}
\def\square{{\mathchoice\sqr74\sqr74\sqr{6.3}3\sqr{3.5}3}}

\def\hcal#1{{\hbox{\cal #1}}}
\def\sst#1{{\scriptscriptstyle #1}}

\def\mpi{{m_\pi}}
\def\mpis{{m^2_\pi}}

\def\mn{{m_\sst{N}}}
\def\mns{{m^2_\sst{N}}}

\def\mro{{m_\rho}}
\def\mros{{m^2_\rho}}

\def\gpnn{{g_{\sst{NN}\pi}}}

\def\Hhat{{\hat H}}

\def\sst#1{{\scriptscriptstyle #1}}
\def\hcal#1{{\hbox{\cal #1}}}
\def\eexp#1{{\hbox{e}^{#1}}}

\def\Hhat{{\hat H}}
\def\That{{\hat T}}

\def\Mhat{{\hat M}}

\def\rohat{{\hat\rho}}

\def\OP{{\hat{\cal O}}}
\def\mn{{m_\sst{N}}}
\def\mns{{m_\sst{N}^2}}

\def\mpi{{m_\pi}}
\def\mpis{{m^2_\pi}}

\def\qv{{\vec q}}
\def\pv{{\vec p}}
\def\ppv{{{\vec p}^{\>\prime}}}
\def\kv{{\vec k}}

\def\kvs{{kv^{\, 2}}}
\def\xv{{\vec x}}
\def\xpv{{{\vec x}^{\>\prime}}}

\def\rv{{\vec r}}

\def\sigv{{\vec\sigma}}
\def\tauv{{\vec\tau}}

\def\gpnn{{g_{\pi\sst{NN}}}}

\def\rbra#1{{\langle#1\parallel}}
\def\rket#1{{\parallel#1\rangle}}

\def\xivz{{\xi_\sst{V}^{(0)}}}

\def\xivtez{{\xi_\sst{V}^{T=0}}}
\def\xivteo{{\xi_\sst{V}^{T=1}}}

\def\FOS{{F_1^{(s)}}}

\def\GES{{G_\sst{E}^{(s)}}}
\def\GMS{{G_\sst{M}^{(s)}}}

\def\mustr{{\mu_s}}

\def\rhostr{{\rho_s}}

\def\GEn{{G_\sst{E}^n}}
\def\GEp{{G_\sst{E}^p}}
\def\GMn{{G_\sst{M}^n}}
\def\GMp{{G_\sst{M}^p}}

\def\GETEZ{{G_\sst{E}^{\sst{T}=0}}}

\def\GMTEZ{{G_\sst{M}^{\sst{T}=0}}}

\def\rbra#1{{\langle#1\parallel}}
\def\rket#1{{\parallel#1\rangle}}

\def\evec{{\vec e}}

\def\mustr{{\mu_s}}

\def\nubar{{\bar\nu}}

\def\kv{{\vec k}}
\def\kvs{{\vec k^2}}

\def\rij{{r_{ij}}}
\def\Rij{{R_{ij}}}
\def\rhij{{{\hat r}_{ij}}}
\def\Rhij{{{\hat R}_{ij}}}
\def\rvij{{{\vec r}_{ij}}}
\def\Rvij{{{\vec R}_{ij}}}
\def\delij{{\vec\Delta_{ij}}}
\def\sigij{{\vec\Sigma_{ij}}}

\def\delv{{\vec\nabla}}

\def\fcztez{{F_{C0}^\sst{T=0}}}
\def\fczs{{F_{C0}^{(s)}}}
\def\beff{{b_{\rm eff}}}
\def\psibar{{\bar\psi}}
\def\kapro{{\kappa_\rho}}
\def\kapom{{\kappa_\omega}}
\def\kapv{{\kappa_\sst{V}}}
\def\gomnn{{g_{\omega\sst{NN}}}}
\def\gronn{{g_{\rho\sst{NN}}}}
\def\gvnn{{g_\sst{VNN}}}

\def\gropitez{{g_{\rho\pi}^\sst{T=0}}}
\def\gropis{{g_{\rho\pi}^{(s)}}}
\def\gropia{{g_{\rho\pi}^{(a)}}}
\def\gropigam{{g_{\rho\pi\gamma}}}
\def\gropiss{{g_{\rho\pi s}}}

\def\GEa{{G_\sst{E}^{(a)}}}
\def\GMa{{G_\sst{M}^{(a)}}}
\def\rohata{{\hat\rho^{(a)}}}
\def\gvdip{{G_\sst{V}^\sst{D}}}
\def\Lv{{\vec L}}
\def\mm{{m_\sst{M}}}
\def\mms{{m^2_\sst{M}}}
\def\lamm{{\Lambda_\sst{M}}}
\def\lamms{{\Lambda^2_\sst{M}}}
\def\gpnns{{g^2_{\pi\sst{NN}}}}
\def\gvnns{{g^2_\sst{VNN}}}
\def\mnc{{m_\sst{N}^3}}
\def\mv{{m_\sst{V}}}
\def\mvs{{m_\sst{V}^2}}
\def\lamsE{{\lambda_\sst{E}^{(s)}}}

\def\PRC#1{{\it Phys. Rev.} {\bf C#1} }
\def\PRD#1{{\it Phys. Rev.} {\bf D#1} }
\def\PRL#1{{\it Phys. Rev. Lett.} {\bf #1} }
\def\NPA#1{{\it Nucl. Phys.} {\bf A#1} }
\def\NPB#1{{\it Nucl. Phys.} {\bf B#1} }
\def\AoP#1{{\it Ann. of Phys.} {\bf #1} }

\def\PLB#1{{\it Phys. Lett.} {\bf B#1} }

\def\ZPC#1{{\it Z. f\"ur Phys.} {\bf C#1} }

\def\overleftrightarrow#1{\vbox{\ialign{##\crcr
	$\leftrightarrow$\crcr\noalign{\kern-1pt\nointerlineskip}
	$\hfil\displaystyle{#1}\hfil$\crcr}}}
\def\delvlr{{\overleftrightarrow\nabla}}
\def\fctez{{F_C^\sst{T=0}}}
\def\fcs{{F_C^{(s)}}}

\hfuzz=50pt

\vsize=7.5in
\hsize=5.6in
\magnification=1200
\tolerance 10000
\input boardmacs

\baselineskip 12pt plus 1pt minus 1pt
\pageno=0
\centerline{\bf Many-Body Currents and the}
\smallskip
\centerline{{\bf Strange-Quark Content of $^4$He}
\footnote{*}{This
work is supported in part by funds
provided by the U. S. Department of Energy (D.O.E.) under contracts
\#DE-AC05-84ER40150 and \#DE-AC02-76ER03069.}
}
\vskip 24pt
\centerline{M. J. Musolf
\footnote{**}{National Science Foundation
Young Investigator}
and R. Schiavilla}
\vskip 12pt
\centerline{\it Department of Physics, Old Dominion University}
\centerline{\it Norfolk, Virginia\ \ 23529\ \ \ U.S.A.}
\centerline{\it and}
\centerline{\it CEBAF Theory Group, MS 12H2}
\centerline{\it Newport News, Virginia\ \ 23606\ \ \ U.S.A.}
\vskip 12pt
\centerline{\it and}
\vskip 12pt
\centerline{T. W. Donnelly}
\vskip 12pt
\centerline{\it Center for Theoretical Physics}
\centerline{\it Laboratory for Nuclear Science}
\centerline{\it and}
\centerline{\it Department of Physics}
\centerline{\it Massachusetts Institute of Technology}
\centerline{\it Cambridge, Massachusetts\ \ 02139\ \ \ U.S.A.}

\vfill
\noindent CEBAF \# TH-94-10\hfill April, 1994
\eject
\baselineskip 16pt plus 2pt minus 2pt
\centerline{\bf ABSTRACT}

	Meson-exchange current (MEC) contributions to the parity-violating (PV)
asymmetry for elastic scattering of polarized electrons from $^4$He are
calculated over a range of momentum transfer using Monte Carlo methods and
a variational $^4$He ground state wavefunction. The results indicate that
MEC's generate a negligible contribution to the asymmetry at low-$|\qv|$,
where a determination of the nucleon's mean square strangeness radius could
be carried out at CEBAF. At larger values of momentum transfer -- beyond the
first diffraction minimum -- two-body corrections from the $\rho$-$\pi$
\lq\lq strangeness charge" operator enter the asymmetry at a potentially
observable level, even in the limit of vanishing strange-quark matrix
elements of the nucleon. For purposes of constraining the nucleon's strangeness
electric form factor, theoretical uncertainties associated with these
MEC contributions do not appear to impose serious limitations.

\vfill
\eject

\noindent {\bf I. Introduction.} One objective of the CEBAF physics program
is to probe the strange-quark \lq\lq content" of the nucleon with
parity-violating (PV) electron scattering. As discussed elsewhere in the
literature [1-7], PV electron scattering at low-to-intermediate energies
is particularly suited to the study of strange-quark vector current matrix
elements, $\bra{H}\sbar\gamma_\mu s\ket{H}$, where $H$ is a hadron. In the
case where the target is a nucleon ($\ket{H}=\ket{p}$ or $\ket{n}$), this
matrix element can  parameterized
by two form factors, $\GES(Q^2)$ and $\GMS(Q^2)$, the
strangeness electric and magnetic form factors, respectively.
Extractions of $\bra{N} \sbar s\ket{N}$, the nucleon's strange-quark scalar
density, from $\pi-N$ scattering [8,9], as well as determinations of
the strange-quark axial vector matrix element, $\bra{N}\sbar\gamma_\mu\gamma_5
s \ket{N}$, from elastic $\nu_\mu p/\nubar_\mu p$ scattering [10-12] and
measurements of the $g_1$ sum [13-15], suggest that
the strange-quark \lq\lq sea" plays a more
important role in the low-energy properties of the nucleon than one might
expect based on the success of valence quark models. Measurements of
$\bra{N}\sbar\gamma_\mu s\ket{N}$ would provide an additional window on the
sea-quark structure of the nucleon. Model estimates of $\GES$ and $\GMS$ at
low-$|Q^2|$ span a wide spectrum in both magnitude and sign [16-21]. It is
therefore of interest to extract the strangeness form factors at a level
needed to distinguish among model calculations and their attendant physical
pictures.

	To this end, use of a proton target would not be sufficient. The
presence of several poorly-constrained form factors in the PV elastic
$^1$H$(\evec, e)$ asymmetry, as well as theoretical uncertainties associated
with axial vector radiative corrections, limit the precision with which
$\GES$ and $\GMS$ could be determined from the proton alone [1,2]. The use of
$A>1$ targets in conjuction with the proton offers the possibility of
imposing more stringent limits on the nucleon's $s$-quark vector current
matrix elements [1,2,22] than could be obtained with a proton target
only. In this regard, the $(J^\pi, T)=(0^+, 0)$ nuclei,
such as $^4$He, constitute an attractive case, since the ground states of
such nuclei can support matrix elements of only one operator -- the isoscalar
Coulomb operator [1,2,22,23]. In the one-body approximation to this operator,
the nuclear wavefunction dependence of the Coulomb matrix elements
effectively cancels out from
the PV asymmetry for such nuclei, leaving only a sensitivity to Standard Model
couplings and single nucleon form factors ({\it e.g.}, $\GES$).  Two
approved CEBAF experiments rely on this feature of
$\alr(0^+,0)$, the PV left-right asymmetry [24,25].
The proper interpretation of $\alr(0^+,0)$ requires that
one understand the importance of many-body corrections to
the one-body asymmetry. Meson-exchange currents (MEC's) constitute one class
of such many-body effects. In previous work [26], we computed MEC
contributions to the $^4$He mean-square \lq\lq strangeness radius", which
generates the leading $s$-quark contribution to $\alr(^4\hbox{He})$ at
low-$|\qv|$. The results of that calculation, peformed with a simple
$^4$He shell model wavefunction and phenomenological two-body correlation
function, indicate that the $^4$He strangeness radius is dominated by
strange-quarks inside the nucleon.

	In the present work, we extend the calculation of Ref.~[26] using
a $^4$He variational wavefunction obtained from realistic
interactions
and computing the asymmetry over the full range of momentum transfer germane
to the future CEBAF experiments. Our results indicate
that the $^4$He strangeness radius is two orders of magnitude more
sensitive to the nucleon's strangeness radius than to two-body contributions.
At the higher $|\qv|$ of experiment [24], the situation is more complex.
Even if the nucleon matrix element $\bra{N}\sbar\gamma_\mu s\ket{N}$ were to
vanish,
the PV asymmetry would still receive a non-negligible contribution from
non-nucleonic $s$-quark matrix elements. In particular, the $\rho-\pi$
strangeness transition charge operator generates nearly a 15\% contribution
to the asymmetry at the kinematics of the experiment [24]. In this case,
an experiment like that
of Ref.~[24] would be significant in two respects. First,
it would be interesting to measure a non-negligible strange-quark matrix
element in a strongly-interacting, non-strange system, regardless
of the dynamical origin
of that matrix element. Second, the only other
observable with significant sensitivity to the $\rho-\pi$ MEC is the $B$ form
factor of the deuteron [27]. If, however,  $\GES$ and $\GMS$ are
non-zero, the level of theoretical {\it uncertainty} associated with the
present MEC calculation does not appear to be large enough to significantly
weaken the possible constraints on $\GES$ which a measurement of
$\alr(^4\hbox{He})$ could provide.

	In the remainder of the paper we provide details of the calculations
leading to these conclusions. Section II gives our formalism, including
expressions for the operators used. In section III, we treat the computation
of the $^4$He matrix elements of these operators, considering first the
simple case of a shell model ground state and subsequently turning to the
Variational Monte Carlo (VMC) approach. In section IV we discuss our results,
including implications for the interpretation of $\alr(^4\hbox{He})$ and
studies of nucleonic strangeness. Technical details may be found in the
Appendix.
\bigskip
\noindent {\bf II. Formalism.} The PV left-right asymmetry for scattering of
polarized electrons from a nuclear target depends on the interference of the
electromagnetic (EM) and PV weak neutral current (NC) amplitudes, $M_\sst{EM}$
and $M^\sst{PV}_\sst{NC}$, as
$$
\alr\approx{2{\rm Re}\  M_\sst{EM}^\ast M_\sst{NC}^\sst{PV}\over
|M_\sst{EM}|^2}
\ \ \ ,\eqno(1)
$$
where $|M_\sst{EM}|>>|M_\sst{NC}^\sst{PV}|$ at low energies. The amplitude
$M_\sst{NC}^\sst{PV}$ is proportional to the sum of two terms,
$$
\ell^\mu_\sst{NC}J_{\mu 5}^\sst{NC}+\ell^{\mu 5}_\sst{NC} J_\mu^\sst{NC}
\ \ \ ,\eqno(2)
$$
where $\ell^{\mu(5)}_\sst{NC}$ is the electron's vector (axial vector) neutral
current and $J_{\mu (5)}^\sst{NC}$ is the nucleon or nuclear matrix element of
the hadronic vector (axial vector)
NC. One may rewrite $\alr$ in terms of quantities which set the
scale of the asymmetry and a ratio of nuclear response functions [1,2]
$$
\alr={G_\mu Q^2\over 2\sqrt{2}\pi\alpha}
{W^\sst{PV}\over F^2}\ \ \ ,\eqno(3)
$$
where $G_\mu$ is the Fermi constant measured in muon-decay, $\alpha$ is the
EM fine structure constant, and $Q^2=\omega^2-q^2$ with $\omega$ and
$q=|\qv |$ being the energy and magnitude of three-momentum transfer to the
target. The response functions appearing in the ratio of Eq.~(3) may be
written as
$$
\eqalignno{F^2&=v_\sst{L}R_\sst{L}+v_\sst{T} R_\sst{T}&(4{\rm a})\cr
	    W^\sst{PV}&=v_\sst{L}W_\sst{AV}^\sst{L}+v_\sst{T}W_\sst{AV}^\sst{T}
		+v_\sst{T'}W_\sst{VA}^\sst{T'}\ \ \ ,&(4{\rm b})\cr}
$$
where $v_\sst{L}$, $v_\sst{T}$ and $v_\sst{T'}$ are leptonic kinematic
factors; $R_\sst{L}$ and $R_\sst{T}$ are the usual longitudinal and
transverse EM response functions; and $W_\sst{AV}^\sst{L,T}$ and
$W_\sst{VA}^\sst{T'}$ are analogous PV response functions involving products
of the hadronic EM and vector NC (\lq\lq $AV$") or axial vector NC (\lq\lq
$VA$") [1,2].

	In this work, we follow the approach taken in Refs.~[1-7] and keep
only the three lightest quarks in the hadronic current. In this case, one
has for the two vector currents
$$
\eqalignno{J_\mu^\sst{EM}&=J_\mu^\sst{EM}(T=1)+J_\mu^\sst{EM}(T=0)&(5{\rm
a})\cr
	J_\mu^\sst{NC}&=\xivteo J_\mu^\sst{NC}(T=1)+\sqrt{3}\xivtez
J_\mu^\sst{EM} (T=0)+\xivz \sbar\gamma_\mu s&(5{\rm b})\cr}
$$
where the $J_\mu^\sst{EM}(T)$ are the isovector $(T=1)$ and isoscalar $(T=0)$
EM currents and the $\xi_\sst{V}^{(a)}$ are couplings determined by the
Standard Model [1,2,22]. A decomposition of $J_{\mu 5}^\sst{NC}$ analogous
to that of Eq.~(5b) but involving the SU(3) octet of axial currents and
$\sbar\gamma_\mu\gamma_5 s$ may also be made [1,2,7]. Since the $^4$He ground
state supports no axial vector matrix element, however, we do not consider
$J_{\mu 5}^\sst{NC}$ further in this work.

	In the limit that  the $^4$He ground state is an eigenstate of isospin,
the \lq\lq hadronic ratio" for this target is
$$
{W^\sst{PV}\over F^2}\> = \> -{1\over 2}\left\{\sqrt{3}\xivtez +
	\xivz{F_{C0}^{(s)}(q)\over F_{C0}^\sst{T=0}(q)}\right\}\ \ \ .
	\eqno(6)
$$
Here, $\sqrt{3}\xivtez=-4\sstw$ and $\xivz=-1$ at tree level in the Standard
Model [1,2,22]. The form factors are given by
$$
\eqalignno{F_{C0}^{(a)}(q)&= \bra{0^+0}\Mhat_{00}^{(a)}(q)\ket{0^+0}&
	(7{\rm a})\cr
&\cr
	\Mhat_{00}^{(a)}(q)&=\int\ d^3x\ j_0(qx)Y_{00}(\Omega_x)\rohat^{(a)}(\xv)&
	(7{\rm b})\cr
&\cr
	&={1\over 4\pi}\int\ d\Omega_q\ Y_{00}(\Omega_q)\rohat^{(a)}(\qv)
	\ \ \ ,&(7{\rm c})\cr}
$$
where $x=|\xv|$ and $\rohat^{(a)}(\xv)$ ($\rohat^{(a)}(\qv)$)
denotes the co-ordinate-space (momentum-space) charge ($\mu=0$) component
of either the isoscalar EM current ($(a)\to T=0$) or strange quark current
($(a)\to (s)$). Matrix elements of the Coulomb operator are simply related
to the elastic charge form factor as
$$
F_C^{(a)}(q)=\bra{0^+0}\rohat(\qv)\ket{0^+0}=2\sqrt{\pi}F_{C0}^{(a)}(q)\ \ \ .
\eqno(7{\rm d})
$$
One observes from Eq.~(6) that were the nuclear matrix
elements of $\Mhat^{(s)}_{00}(q)$ to vanish, the asymmetry would be
nominally independent of the details of the nuclear
wavefunction.\footnote{\dag}{Apart from contributions from nuclear dispersion
corrections; see, {\it e.g.} Refs.~[1,2,22].} The reason is that (i) in the
absence
of strangeness, the hadronic isoscalar EM and isoscalar NC currents are
identical, up to the overall electroweak coupling, $\sqrt{3}\xivtez$, (ii)
isovector matrix elements vanish if the $^4$He ground state is assumed
to be a pure $T=0$ state, and
(iii) a spin-0 ground state cannot support axial vector
matrix elements.
\medskip
\noindent\undertext{One-body operators}
\medskip

	Expressions for the one-body charge operators may be obtained starting
from Lorentz-covariant forms of the single-nucleon vector current matrix
element:
$$
\bra{N(p')} V_\mu(0)\ket{N(p)} = {\bar U}(p')\left[F_1(Q^2)\gamma_\mu +
	{iF_2(Q^2)\over 2\mn}\sigma_{\mu\nu}Q^\nu\right]U(p)\ \ \ ,\eqno(8)
$$
where $F_1$ and $F_2$ are the standard Dirac and Pauli form factors of
the nucleon, $U(p)$ and $U(p')$ are nucleon spinors corresponding to nucleon
states $\ket{N(p)}$ and $\ket{N(p')}$, respectively, and $V_\mu(\xv)$ is
any one of the vector currents of interest (isoscalar EM or strangeness).
Expanding the right side of Eq.~(8) in powers of $p/\mn$, transforming to
co-ordinate space, and summing over all nucleons gives for the $\mu=0$
component
$$
\rohata(\qv)^{[1]}=\sum_{k=1}^{A}\eexp{i\qv\cdot\xv_k}\left[{
	\GEa(\tau)\over\sqrt{1+\tau}}
-{i\over 8\mns}\left\{\GEa(\tau)-2\GMa(\tau)\right\}\sigv_k
	\cdot\qv\times{\vec P_k}\right]\ \ \ ,\eqno(9)
$$
where $\tau\equiv -Q^2/4\mns = q^2/4\mns$ for elastic scattering
in the Breit frame, $\vec P_k =
\pv_k+\ppv_k$, and
$$
\eqalignno{\GEa&=F_1^{(a)}-\tau F_2^{(a)}&(10{\rm a})\cr
	   \GMa&=F_1^{(a)}+F_2^{(a)}&(10{\rm b})\cr}
$$
are the Sachs electric (10a) and magnetic (10b) form factors [28]. In
arriving at the expression in Eq.~(9), we have used the spinor normalization
of Ref.~[29]. Had we followed the convention of Ref.~[30],
the charge operator would have contained an additional term $\Delta_k$ inside
the square brackets given by
$$
\Delta_k= {1\over 4\mns}\left(p^2_k+{p^\prime}_k^2\right)(1+\tau)^{-1}\left[
	\GEa(\tau)+\tau\GMa(\tau)\right]\ \ \ .\eqno(11)
$$

	Following the convention in Refs.~[1-4], we parameterize the
momentum-dependence of the one-body form factors as
$$
\eqalignno{\GEp(\tau)&=\gvdip(\tau)&(12{\rm a})\cr
	   \GMp(\tau)&=\mu_p\gvdip(\tau)&(12{\rm b})\cr}
$$
$$
\eqalignno{\GEn(\tau)&=-\mu_n\tau\gvdip(\tau)\xi_n(\tau)&(12{\rm c})\cr
	   \GMn(\tau)&=\mu_n\gvdip(\tau)&(12{\rm d})\cr}
$$
$$
\eqalignno{\GES(\tau)&=\rhostr\tau\gvdip(\tau)\xi_s(\tau)&(12{\rm e})\cr
	   \GMS(\tau)&=\mustr\gvdip(\tau)&(12{\rm f})\cr}
$$
with
$$
\eqalignno{\gvdip(\tau)&=(1+\lambda_\sst{V}^\sst{D}\tau)^{-2}&(13{\rm a})\cr
	   \xi_n&=(1+\lambda_n\tau)^{-1}&(13{\rm b})\cr
	   \xi_s&=(1+\lamsE\tau)^{-1}&(13{\rm c})\cr}
$$
and
$$
\eqalignno{G_\sst{E,M}^\sst{T=0}&={1\over 2}\left[G_\sst{E,M}^p+G_\sst{E,M}^n
	\right]&(14{\rm a})\cr
	   G_\sst{E,M}^\sst{T=1}&={1\over 2}\left[G_\sst{E,M}^p-G_\sst{E,M}^n
	\right]\ \ \ .&(14{\rm b})\cr}
$$
Numerically, one has $\mu_p=2.79$, $\mu_n=-1.91$, $\lambda_\sst{V}^\sst{D}=
4.97$, and $\lambda_n=5.6$. The rationale for adopting this parameterization
is discussed more fully in Refs.~[1,2]. The parameters $\mustr$ and
$\rhostr$, which define the strangeness magnetic moment and strangeness
radius, respectively, as well as $\lamsE$ which governs the next-to-leading
$Q^2$ behavior of $\GES$, are presently unknown. One goal of the SAMPLE
experiment [6] and up-coming CEBAF experiments [24,25,31] is to place limits
on these parameters.

	The one-body contribution to the Coulomb multipole operator, obtained
by substituting the expression for the charge operator of Eq.~(9) into
Eq.~(7b), is
$$
\eqalignno{\Mhat_{00}^{(a)}(q)^{[1]}&={1\over 2\sqrt{\pi}}\sum_{k=1}^A
	\biggl\{{\GEa(\tau)\over\sqrt{1+\tau}} j_0(qx_k)&(15)\cr
	&+\left[\GEa(\tau)-2\GMa(\tau)\right]{q\over 2\mn}
	{j_1(qx_k)\over\mn x_k}
	\sigv_k\cdot\Lv_k\biggr\} \ \ \ ,&\cr}
$$
where we have assumed $q_0=0$ so that $q^2=4\mns\tau$ and where $\Lv_k$
is the orbital angular momentum of the $k$-th nucleon. Note that in the limit
that the $^4$He ground state consists of nucleons in S-states only, the
spin-orbit operator in $\Mhat_{00}^{(a)}(q)^{[1]}$ will not contribute to
$F_{C0}^{(a)}(q)$. In this case, the Coulomb matrix elements for the
isoscalar EM and strangeness charge operators are identical, apart from the
single nucleon form factors, rendering their ratio independent of nuclear
structure:
$$
{F_{C0}^{(s)}(q)^{[1]}\over F_{C0}^\sst{T=0}(q)^{[1]}}\bigg\vert_{\rm S-waves}
\longrightarrow {\GES(\tau)\over\GETEZ(\tau)}\ \ \ .\eqno(16)
$$
However, the presence of a significant D-wave
component (the associated probability is about 16\% for
the variational $^4$He wavefunction discussed below)
implies some level of structure-dependence
in the one-body form factor ratio of Eq.~(16). For most values of momentum
transfer, the magnitude of this structure-dependence is negligible (see
the spin-orbit contributions in Figs. 2b and 3b).

\goodbreak
\bigskip
\noindent\undertext{Two-body operators}
\medskip

	In the one boson-exchange (OBE) approximation, the leading two-body
MEC corrections to the one-body result of Eq.~(16) are generated by the
processes in Fig. 1. The $\pi$-exchange and vector meson-exchange
\lq\lq pair currents" (Figs. 1a,b) are familiar from previous work on MEC's
[32-36], as is the pseudoscalar-vector meson \lq\lq transition current" of
Fig. 1c. In each case, the isoscalar EM and strangeness two-body currents
have the same structure, apart from the form factors appearing at the
$N\bar N$ creation/annihilation vertex and $V-\pi$ transition vertex. The
$^4$He elastic from factors receive no contribution from processes in which
a virtual $\gamma$ or $Z^0$ couples to an exchanged pseudoscalar or vector
meson. The reason is that matrix elements
of the form $\bra{M}V_\mu(0)\ket{M'}$
must vanish in order to respect G-parity invariance when $\ket{M'}$ and
$\ket{M}$ are identical meson states (apart from momenta) and when $V_\mu$
is either $J_\mu^\sst{EM}(T=0)$ or $\sbar\gamma_\mu s$. Moreover, one
has no contribution from an $\omega-\pi$ transition current since currents
which are strong isoscalar operators cannot induce such an isospin-changing
transition. We have not included contributions from isobar currents, since the
lightest nucleon resonance accessible with an isoscalar current is the
$N(1440)$.
We assume contributions from the associated current are suppressed by the large
mass difference between this state and the nucleon.

	We derive two-body charge operators by computing the covariant
momentum-space Feynman amplitudes associated with the diagrams in Fig. 1,
peforming the standard non-relativistic reduction, and transforming to
co-ordinate space. We take the meson-nucleon couplings from the conventional
low-energy effective Lagrangians:

$$
\eqalignno{{\cal L}_{NN\pi}&={\gpnn\over 2\mn}\psibar_N(x)\notcder{\vec\pi}(x)
	\cdot\tauv\psi_N(x)&(17{\rm a})\cr
&\cr
	{\cal L}_{NN\rho}&=\gronn\psibar_N(x)\left[\gamma^\mu+{\kapro\over
	2\mn}\sigma^{\mu\nu}\partial_\nu\right]{\vec\rho}_\mu\cdot\tauv
	\psi_N(x)&(17{\rm b})\cr
&\cr
	{\cal L}_{NN\omega}&=\gomnn\psibar_N(x)\left[\gamma^\mu+{\kapom\over
	2\mn}\sigma^{\mu\nu}\partial_\nu\right]\omega_\mu
	\psi_N(x)&(17{\rm c})\cr }
$$
and
$$
D^\mu = \partial^\mu+ie{\hat Q}_\sst{EM} A^\mu+ig{\hat Q}_\sst{W} Z^\mu\ \ \ ,
\eqno(17{\rm d})
$$
where $\psi_N$ is a nucleon field, $\pi^a$, $\rho^a_\mu$, and $\omega_\mu$
are the pion, rho-meson, and omega-meson fields, respectively, \lq\lq $a$"
is an isospin index, $g$ is the semi-weak coupling, $\hat Q_\sst{EM}$ and
$\hat Q_\sst{W}$ are the EM and weak NC charge operators, and $A^\mu$ and
$Z^\mu$ are the photon and $Z^0$ fields, respectively. We take the couplings
appearing in Eq.~(17) to have the values $\gpnn=13.6$, $\gronn=2.6$,
$\gomnn=14.6$, $\kapro=6.6$, and $\kapom=-0.12$ [37]. Momentum-space
matrix elements of the operators in Eq.~(17) have the same structure as
the effective Lagrangians, but with the nucleon fields replaced by plane
wave spinors, the derivatives replaced by $i k^\mu$, where $k^\mu$ is the
momentum of the outgoing meson, and the vector boson fields replaced by the
corresponding polarization vectors, $\varepsilon_\mu$. For the $\rho-\pi$
transition current matrix element one has

$$
\bra{\pi^b(k_2)}V_\mu^{(a)}(0)\ket{\rho^c(k_1,\varepsilon)}=-{\gropia(Q^2)\over
\mro}\delta_{bc}\epsilon_{\mu\nu\alpha\beta}k_1^\nu k_2^\alpha\varepsilon^\beta
\ \ \ ,\eqno(18)
$$
where as usual \lq\lq $a$" denotes either the EM or strange-quark
current [38].
In the case of the former, the value of the transition form factor
at the photon point is known to be $\gropitez(Q^2=0)\equiv\gropigam=0.56$
[39], while the $Q^2$-dependence may be modelled using $\omega$-pole
dominance:
$$
\gropitez(Q^2)=\gropigam\left(1-Q^2/m_\omega^2\right)^{-1}\ \ \ .
\eqno(19)
$$
In the case where $V_\mu^{(a)}=\sbar\gamma_\mu s$, one may follow a similar
approach and assume $\phi$-meson dominance, which is reasonable since the
$\phi$ is almost pure $s\sbar$:
$$
\gropis(Q^2)=\gropiss\left(1-Q^2/m_\phi^2\right)^{-1}\ \ \ .\eqno(20)
$$
The measured rates for $\phi\to\rho\pi$ and $\phi\to\ell^+\ell^-$
($\ell$ is a charged lepton) can be used
to estimate the value of this form factor at $Q^2=0$ to  be $|\gropiss|=
0.26$ [40].

	Before proceeding, we touch on one issue associated with the vector
meson pair current operators. These operators are derived by keeping only the
negative-energy pole of the nucleon propagator, as shown in
Fig. 1. The resulting two-body nuclear matrix element is thus distinct
from the matrix element containing the positive energy pole, which contributes
via the full nuclear Green's function in time-ordered perturbation theory:
$$
\sum_n\left[ {\bra{f} J_\mu\ket{n}\bra{n}\Hhat_\sst{NUC}\ket{i}\over E_i-E_n+
i\varepsilon}+{\bra{f} \Hhat_\sst{NUC}\ket{n}\bra{n} J_\mu\ket{i}\over E_i-E_n+
i\varepsilon}\right] \ \ \ .\eqno(21)
$$
Here, $\Hhat_\sst{NUC}$ is the full nuclear Hamiltonian and $(i,f,n)$ denote
initial, final, and intermediate nuclear states, respectively. Following
this prescription leads one to a two-body pair-current operator having the
same form as given in Ref.~[36]. It has been argued, however,
that one must include an additional retardation contribution arising from
the positive-energy pole in the nucleon propagator whose residue contains
a dependence on the energy transfer between the two nucleons. Inclusion of
this additional term results in the form for the pair-current charge
operator given in Refs.~[32,33]. Rather than attempting to choose between
these two approaches, we compute $F_{C0}^{(s)}$ in two ways -- once
using each of these two prescriptions -- in order to determine the impact
of this choice. As we note in Section IV, the vector-meson exchange
contributions to the $^4$He form factors are sufficiently small in comparison
with other contributions that the impact of this choice in the value for
$F_{C0}^{(s)}$ is insignificant.

	The momentum-space charge operators for the pair currents are
$$
\eqalignno{\rohat(\pv_1, \ppv_1, \pv_2, \ppv_2;\qv)^{[2]}_{\rm pionic}
&=(2\pi)^3\delta
(\kv_1+\kv_2-\qv)\left[{\gpnns\over 4\mnc}\right] F_1^{(i)}(\tau)\tauv_1
\, \cdot\tauv_2&(22{\rm a})\cr
&\times\left\{ { {1}\over{\kv^2_2+\mpis} }\sigv_1\cdot\qv
\, \sigv_2\cdot\kv_2+(1\leftrightarrow 2)\right\}\ \ \ ,&\cr}
$$
where $\kv_i=\ppv_i-\pv_i$, $i=1,2$, and
$$
\eqalignno{\rohat(\pv_1, \ppv_1, \pv_2, \ppv_2;\qv)^{[2]\ \rm (a)}_{\rm V-pair}
&=(2\pi)^3\delta
(\kv_1+\kv_2-\qv)\left[{\gvnns\over 4\mnc}\right]G_\sst{M}^{(i)}(\tau)
\That_\sst{V}(1,2)&(22{\rm b})\cr
&\times\Biggl\{ {1\over\kv_2^2+\mvs} \biggl[(1+
\kappa_\sst{V})\left(\qv\cdot\kv_2+\sigv_1\times\qv\cdot\sigv_2\times\kv_2
\right)&\cr
&-i\sigv_1\times\qv\cdot(\pv_2+\ppv_2)\biggr]+(1\leftrightarrow 2)\Biggr\}&\cr}
$$
excluding the retardation correction or
$$
\eqalignno{\rohat(&\pv_1, \ppv_1, \pv_2, \ppv_2;\qv)^{[2]\ \rm (b)}_{\rm
V-pair}
=(2\pi)^3\delta
(\kv_1+\kv_2-\qv)\left[{\gvnns\over 4\mnc}\right]
\That_\sst{V}(1,2)&(22{\rm c})\cr
&\times\Biggl\{{1\over\kv_2^2+\mvs}\biggl[\left((1+
\kappa_\sst{V})^2F_1^{(i)}(\tau)+(1+\kappa_\sst{V})F_2^{(i)}(\tau)\right)
\sigv_1\times\qv\cdot\sigv_2\times\kv_2&\cr
&+G_\sst{M}^{(i)}(\tau)\left((1+\kappa_\sst{V})\qv\cdot\kv_2-i\sigv_1\times
\qv\cdot(\pv_2+\ppv_2)\right)\biggr]+(1\leftrightarrow 2)\Biggr\}&\cr}
$$
including the retardation term, with
$$
\That_\sst{V}(1,2)=\cases{\tauv_1\cdot\tauv_2\ ,&$V=\rho$\cr
1\ ,&$V=\omega$\cr}\eqno(23)
$$
and where \lq\lq $(i)$'' indicates either the isoscalar EM or strange-quark
charge operators. We have not included the isovector parts of the charge
operators. For the transition operators, one has
$$
\eqalignno{\rohat(\pv_1, \ppv_1, \pv_2, \ppv_2; \qv)^{[2]}_{\rho\pi}
&=i(2\pi)^3
\delta(\kv_1+\kv_2-\qv)\left[{\gpnn\gronn\gropis(Q^2)\over 4\mro\mns}\right]
\tauv_1\cdot\tauv_2&(24)\cr
&\times\Biggl\{ { {1}\over { (\kv_1^2+\mpis)
(\kv_2^2+\mros) } }\sigv_1\cdot\kv_1\biggl[(\pv_1+\ppv_1)
\cdot(\kv_1\times\kv_2)&\cr
&-i(1+\kapro)(\kv_1\times\kv_2)\cdot(\kv_2\times\sigv_2)\biggr]+(1
\leftrightarrow 2)\Biggr\}\ \ \ .&\cr}
$$

	Expressions for the co-ordinate space forms of the two-body
charge operators, $\rohat(\xv_1, \xpv_1, \xv_2, \xpv_2; \qv)^{[2]}$,
as well as for their Coulomb multipole projections, $\Mhat_{00}(\qv)^{[2]}$,
are somewhat involved and may be found in the Appendix. For purposes of
discussion, it is useful to consider the leading-$q$ behavior of
the two-body Coulomb operators (shown in Eqs. (A.10) of the Appendix),
since their matrix elements contribute to the $^4$He EM and
strangeness radii. From the low-$q$ expressions for the two-body
Coulomb operators, we observe that they vanish at least
as rapidly as $q^2$ for small $q$. The operators must vanish
at $q^2=0$, since the two-body operators cannot change the overall charge
(EM or strangeness) of the $^4$He nucleus. In the case of strangeness,
the entire nuclear form factor $F_{C0}^{(s)}$ must vanish at $q^2=0$,
since the nucleus has no net strangeness. Thus, in analogy with
the single nucleon case, we define a nuclear strangeness radius as
$$
\rhostr[{\rm nuc}]=2\sqrt{\pi}{d F_{C0}^{(s)}\over d\tau}
\Bigg\vert_{\tau=0}\ \ \ .\eqno(25)
$$
Under this definition, $\rhostr[{\rm nuc}]=A\rhostr$ in the one-body limit
neglecting the spin-orbit contribution.  From the expressions
in Eq.~(A.10), we note that
the pionic operator (Eq. (A.10a))
contributes to $F_{C0}^{(s)}$ at ${\cal O}(q^4)$, since this operator is
proportional to  $q^2 F_1^{(a)}$ and since $F_1^{(a)}$
vanishes as $q^2$ for small $q$.
Consequently, the longest-range MEC does not contribute to the nuclear
strangeness radius. For the same reason, the retardation correction to the
vector meson pair current operator (Eq. (22c) and Ref. [33])
also does not contribute to
$\rhostr[{\rm nuc}]$, since this correction is proportional to $\tau F_1^{(s)}(
\tau)$. As a result, the low-$q$ behavior of the vector meson contribution to
$\fczs$ is independent of the choice of
approach discussed above. This choice takes on relevance only at larger
values of momentum-transfer, where the terms proportional to $\tau F_1^{(s)}
(\tau)$ are non-negligible.

\medskip
\noindent{\bf III. Calculation of $^4$He Matrix Elements.} Although the object
of this paper is to report on a calculation of $\fcztez$ and $\fczs$ using
state
of the art wavefunctions, we first summarize a simpler calculation of the
$^4$He strangeness radius using a shell model ground state with harmonic
oscillator wavefunctions. This simpler treatment allows for an analytical
computation and serves to guide one's intuition when interpreting results
obtained with more sophisticated methods. The results of the shell model
calculation were reported previously [26], and we provide more details
in the Appendix of the present paper.

\medskip
\noindent\undertext{Shell Model Calculation}
\medskip

	In the simplest shell model description of $^4$He, the ground
state consists of a single configuration: four nucleons in the $1s_{1/2}$
state. Numerical results using more realistic wavefunctions, such as
the variational wavefunction described below, suggest that the level
of configuration mixing is at least 15\% . Within the S-state approximation,
we compute the leading-$q$ behavior of $\fczs$ using harmonic oscillator
single-particle wavefunctions with an oscillator parameter $b=1.2$ fm,
obtained from fits to the data on $\fcztez$ [2]. Analytic
expressions for the nuclear matrix elements appear in the Appendix, and our
results give
$$
\eqalignno{\fczs(\tau\to 0)&={1\over 2\sqrt{\pi}}\tau\rhostr[^4{\rm He}]
&(26)\cr
&\cr
&=\tau\left(\lambda_1\rhostr+\lambda_{2\rm a}\mustr+\lambda_{2\rm b}
\gropiss\right)
\ \ \ ,&\cr}
$$
where the terms containing $\lambda_1$ and $\lambda_{2\rm a,b}$ give the
one- and two-body contributions, respectively. The one-body term is
nuclear structure-independent, since the leading $q$-dependence of
the one-body strangeness Coulomb operator is given by $\GES$ times
an operator which counts the number of nucleons (see Eq. (15)). The
two-body term $\lambda_{2\rm a}\mustr$ arises from the vector meson
pair currents, while the term $\lambda_{2\rm b}\gropiss$ is generated
by the $\rho$-$\pi$ transition current. Numerically, in the limit of
point meson-nucleon vertices ($\Lambda_\sst{M}\to\infty$), we obtain
$\lambda_1\approx 1.13$, $\lambda_{2\rm a}\approx -0.05$, and
$\lambda_{2\rm b}\approx -0.02$ after including a phenomenological
NN anti-correlation function in the two-body matrix elements. We expect
that the values of the $\lambda_{2\rm a,b}$
for finite $\Lambda_\sst{M}$ should be smaller in magnitude than
those quoted, which we take to give an upper bound on the scale of
two-body contributions. These results imply, then,
that $\rhostr[^4{\rm He}]$ is at least a factor of 20
more sensitive to the nucleon's strangeness radius than to two-body
strangeness currents.

	We note in passing that had we not accounted for short
range NN repulsion, the vector meson contribution would have been
a factor two larger in magnitude and the $\rho$-$\pi$ term matrix
element would have been a factor of ten larger. The reason for the
large suppression of the $\rho$-$\pi$ term due to short-range
repulsion can be seen from the structure of the momentum-space
$\rho$-$\pi$ charge operator in Eq.~(24). At leading-order in $q$,
the Coulomb projection of this operator has the form
$$
\eqalignno{q^2\OP\left[{1\over (\kvs_1+\mpis)(\kv_1^2+\mros)}\right]
&+\hcal{O}(q^4)&(27)\cr
&={q^2\over(\mros-\mpis)}\OP\left[{1\over \kvs_1+\mpis}-
{1\over \kvs_1+\mros}\right]+\hcal{O}(q^4)\ \ \ ,&\cr}
$$
where $\OP$ is an operator dependent on $\sigv_{1,2}$ and $\kv_1$. Nuclear
matrix elements of the full operator in Eq.~(27) thus depend  on the
difference of matrix elements of two operators, $\hat A(\mpi)$
and $\hat A(\mro)$,
whose ranges are set by $\mpi$ and $\mro$, respectively. In the absence of
short-range anti-correlations,
one has $2\rbra{\rm g.s.}\hat A(\mpi)\rket{\rm g.s.}\approx
\rbra{\rm g.s.}\hat A(\mro)\rket{g.s.}$. The impact of short range repulsion
is to reduce the $\rho$-meson term $\rbra{\rm g.s.}\hat A(\mro)\rket{g.s.}$
by about a factor of two, while leaving the matrix element of the pionic
operator, whose range is much larger than the radius
of the repulsive core, relatively unchanged.
Consequently, the degree of cancellation between the two pieces is
greatly enhanced, leading to the factor of ten reduction in $\lambda_{2\rm b}$,
as compared with the less significant impact on the magnitude of the
purely vector meson matrix elements, $\lambda_{2\rm a}$.

\medskip
\noindent\undertext{Variational Monte Carlo Calculation}
\medskip
The $^4$He variational wavefunction
used in the present work is obtained by minimizing
a realistic Hamiltonian with the Argonne $v_{14}$ two-nucleon [41] and
Urbana-VIII three-nucleon [42] interaction models.  It has the symmetrized
product form given by [42]:
$$
|\Psi> = \bigl[ 1 + \sum_{i<j<k} U^{TNI}_{ijk} \bigr]
\bigl[ S \prod_{i<j} ( 1+U_{ij} ) \bigr] |\Psi_{\rm J}> \>\>.\eqno(28)
$$
\noindent Here $S$ is the symmetrizer, and $|\Psi_{\rm J}>$ is a Jastrow
wavefunction
$$
|\Psi_{\rm J}> = \bigl[ \prod_{i<j} f^c(r_{ij})\bigr] A | \uparrow p \downarrow
p
 \uparrow n \downarrow n > \>\> ,\eqno(29)
$$
\noindent where $A$ is the antisymmetrizer acting on the spin-isospin
states of the four nucleons.  The two-body correlation
operator $U_{ij}$ is taken to be
$$
U_{ij} =  \sum_{p=\tau, \sigma, \sigma \tau, t, t\tau} u^p(r_{ij}) O^p_{ij}
\eqno(30)
$$
\noindent with
$$
O^p_{ij}= \vec \tau_i \cdot \vec \tau_j \>,\>
          \vec \sigma_i \cdot \vec \sigma_j \>,\>
          \vec \sigma_i \cdot \vec \sigma_j \vec \tau_i \cdot \vec \tau_j \>,\>
          S_{ij} \>,\> S_{ij} \vec \tau_i \cdot \vec \tau_j \>\>;\>\>
p=\tau,\sigma,
          \sigma \tau, t , t\tau \>\>.\eqno(31)
$$
\noindent The three-body correlation operator $U^{TNI}_{ijk}$ is simply
related to the three-nucleon interaction present in the Hamiltonian, and has a
correspondingly complex operator dependence.  The
correlation functions $f^c(r)$ and $u^p(r)$ as well as the additional
parameters
present in $U^{TNI}_{ijk}$ are determined
variationally with the methods discussed
in detail in ref. [42].
\par The $^4$He binding energy and charge radius calculated with the above
wavefunction have errors of $\simeq$ 4\% when compared to exact Green's
Function
Monte Carlo (GFMC) results for
the same Hamiltonian [42,43] (we note that
the GFMC results reproduce the empirical
values).  This wavefunction also produces a charge form factor that is
in good agreement with the exact GFMC predictions and the experimental data
over a wide range of momentum transfers [42].  Because of the
relatively strong tensor component in the Argonne $v_{14}$ the D-state
probability has the rather large value of 16\%.
\par The charge and strangeness form factors are given by the
expectation values
$$
F_C^{(a)} = 2\sqrt{\pi} F_{C0}^{(a)} (q) =
<\Psi;\vec q \, | \hat \rho ^{(a)}(\vec q \,) |\Psi>\>\>\>, \eqno(32)
$$
\noindent where $|\Psi; \vec q>$ denotes the ground state wavefunction
recoiling with momentum $\vec q$, and $\hat \rho^{(a)}(\vec q\,)$ are
the $r$-space representations of the charge and strangeness operators
listed in the appendix.  The above expectation value is computed, without
any approximation, by Monte Carlo integration.  The wavefunction is written
as a vector in the spin-isospin space of the A-nucleons for any given
spatial configuration $\vec R \equiv ({\vec r_1,...,\vec r_{\rm A}})$.  For
the given $\vec R$, we calculate the state vector
$\hat \rho^{(a)}(\vec q \,)|\Psi>$ by performing exactly the spin-isospin
algebra with the methods developed in refs. [32,44].  The momentum-dependent
terms in $\hat \rho^{(a)}$ are calculated numerically; for example,
$$
\nabla_{i,\alpha}\Psi(\vec R)={1\over 2 \delta_{i,\alpha}}\lbrack
\Psi(\vec R+\delta_{i,\alpha})-\Psi(\vec R-\delta_{i,\alpha}) \rbrack \>\>,
\eqno(33)
$$
where $\delta_{i,\alpha}$ is a small increment in the $r_{i,\alpha}$
component of $\vec R$.  The $\vec R$-integration is
carried out with Monte Carlo techniques by sampling
a large set of $\vec R$ configurations with the Metropolis algorithm.
\par The two-body pion and $\rho$-meson operators have been constructed
from the Argonne $v_{14}$ following the method outlined in ref. [44].  This
implies replacing the
propagators in eqs. (22a-c) by the Fourier transforms
$v^{\sigma\tau}(k)$ and $v^{t\tau}(k)$ of the isospin dependent spin-spin
and tensor components of the interaction model as
$$
\eqalignno{{g_{\pi\sst{NN}}^2 \over {4 m_N^2}} {1 \over { k^2+m^2_{\pi} }}
&\rightarrow V_{\pi}(k)=  2v^{t\tau}(k)-v^{\sigma\tau}(k)&(34{\rm a})\cr
-{g_{\rho\sst{NN}}^2 (1+\kappa_\rho)^2 \over {4\mns}}
 {1\over { k^2+m^2_{\rho} }}
&\rightarrow V_{\rho}(k)=  v^{t\tau}(k)+v^{\sigma\tau}(k)&(34{\rm b})\cr}
$$
\noindent The replacements eq.(34) are the ones required for the construction
of a two-body electromagnetic current operator that satisfies the
continuity equation with the interaction model [44].  We here apply this
replacement to the pair current EM and strangeness charge operators as the
generalized propagators constructed in this way are then consistent with
the short-range behavior of the corresponding interaction components.
This short-range behavior is determined phenomenologically by fitting
NN elastic scattering data.  An additional justification for using the
construction eq.(34) is that it has been shown to lead to predictions for the
charge and magnetic form factors of the trinucleons [32,42,44], and
threshold electrodisintegration of the deuteron [45] that are in reasonably
good agreement with the empirical data. The $\omega$-meson propagator in
the corresponding pair current, Eqs. (22b,c), and the $\rho$- and
$\pi$-meson propagators in the transition current, Eq.(23), are
modified by the inclusion of monopole meson-nucleon form factors
$$
F_\sst{NNM}(k^2)={\lamms-\mms\over k^2+\lamms}\ \ \ ,\eqno(35)
$$
where $M$ is the exchanged meson of mass $\mm$ and $\lamm$ is a cut-off
parameter.
We use the values $\Lambda_{NN\omega}=\Lambda_{NN\rho}=\Lambda_{NN\pi}=2$ GeV,
as
obtained in boson exchange interaction models [46].  It should be emphasized
that the contributions due to the vector meson pair currents are not
significant in the momentum transfer range of interest here.  Furthermore, we
note that in evaluating the contributions due to the vector meson
pair currents that include the retardation correction, the non-local
terms in eq.(22c), namely those proportional to $\vec p+\vec p^{\, \prime}$,
have
been neglected.  This is justified for the $\rho$-meson pair current, since the
non-local
contribution is suppressed by a factor $(1+\kappa_{\rho})^2$ ($\kappa_{\rho}=
6.6$) with respect to the leading term proportional to $F_1^{(s)}(\tau)$.  This
approximation, however, is questionable for the $\omega$-meson pair current,
since in this
case the tensor coupling is small, $\kappa_{\omega}=-0.12$.

\bigskip
\noindent{\bf IV. Results and Discussion.} The results of our
VMC calculation are displayed
in Figs. 2-6. In computing various contributions to $\fcztez$
and $\fczs$, we have employed a value of $\lamsE=\lamn=5.6$ to serve as a
point of comparison, although $\lamsE$ is essentially a free parameter
characterizing the next-to-leading $Q^2$-dependence of $\GES$ and is to be
constrained by experiment.

Assuming the values $\rho_s=-2.12$ and $\mu_s=-0.2$ for the
strangeness radius and magnetic moment of the nucleon, we find
that the relativistic Darwin-Foldy and spin-orbit corrections
to the single nucleon operator, and the two-body contributions
associated with pseudoscalar and vector meson exchanges as well as
the $\rho \pi $ transition current lead to about 0.5\%
decrease (increase in magnitude) of $\rho_s[^4{\rm He}]$, a negligible effect.

	Results for $\fctez=2\sqrt{\pi}\fcztez$ and
$\fcs=2\sqrt{\pi}\fczs$ over a range of momentum-transfer
are shown in Figs. 2 and 3. Panels 2(a) and 3(a) give the
full form factor resulting from the one- and two-body currents as well
as in the impulse approximation (IA) for comparison. Panels 2(b) and
3(b) display individual contributions from the various one- and two-body
terms. As indicated by the plot in Fig. 2(a) and as noted in previous
work [32], the inclusion of MEC's significantly improves the degree of
agreement with the data on $\fctez$ over a wide range of $q$ as compared with
the IA form factor. The difference in
behavior between $\fctez$ and $\fcs$ at
low-$q$ is dictated by the different values of the corresponding nuclear
charges: $\fctez(0)=A\GETEZ(0)=2$ and $\fcs(0)=A\GES(0) = 0$.
At larger values of $q$, the nuclear EM and strangeness
form factors manifest similar structures, having their first diffraction
mimina and subsequent maxima at essentially the same values of
momentum transfer. Since the various contributions to $\fctez(q)$ are
discussed elsewhere [32], we focus on $\fcs(q)$. At low momentum transfer,
the nuclear strangeness form factor is dominated by the single nucleon
contribution proportional to $\GES$. In this regime, the largest corrections
arise from the spin-orbit and $\rho-\pi$ transition currents. At moderate
values of momentum transfer
($q\rapp 2$ fm), the largest corrections are due to the pionic
pair and $\rho-\pi$ transition currents. In arriving at the results
shown in this figure, we have assumed essentially the Jaffe value for
the nucleon's strangeness radius ($\rhostr\approx -2)$ and a value of
$\mustr= -0.2$. Under this assumption of what would be a large
magnitude for $\rhostr$, the one-body, pionic, and $\rho-\pi$ transition
contributions are of the same order of magnitude at the kinematics of the
approved CEBAF experiment [24] ($q=3.93$ fm$^{-1}$). At this point, the
$\rho$-$\pi$ contribution makes up about 20\% of the total $\fcs$. Were
we to employ, instead, the results of the kaon-loop estimates of the
strangeness parameters, $\rhostr\approx 0.4$ and $\mustr\approx -0.3$
[20], the magnitude of $\fcs$ would be an order of
magnitude smaller and would be dominated by the $\rho$-$\pi$ contribution.

	We emphasize that the relative importance of the pionic operator
is highly dependent on one's model for the one-body strangeness form
factor, $\FOS$, which enters the two-body operator multiplicatively.
In this case, the scale of the two-body operator is set by the Dirac one-body
strangeness radius,
$$
\rho_s^{\rm Dirac}={d\FOS\over d\tau}\bigg\vert_{\tau=0}
=\rho_s^{\rm Sachs}+\mustr\ \ \ .\eqno(36)
$$
In arriving at the results displayed in Fig. 3, we used essentially
the pole model value [16] for $\rho_s^{\rm Dirac}\approx -2.4$, but not
the $Q^2$-dependence for $\FOS$, since the latter is un-realistically
gentle in light of simple quark counting arguments. Had we used, instead,
the results of the kaon loop estimate of Ref.~[20], the magnitude
of the pionic contribution would have been a factor of 20 smaller
than the contribution shown in Fig. 3, and the sign would have been
opposite. Similarly, the one-body IA contribution would be reduced by
at least a factor of four in magnitude and its sign would also have been
opposite than what appears in Fig. 3. In this case, the $\rho-\pi$
transition current would generate the dominant contribution to $\fcs$
at the kinematics of the approved CEBAF experiment [24], while the single
nucleon strangeness radius would still govern the low-$q$ behavior of
the nuclear strangeness form factor.

	By way of comparison, we note that the vector meson pair current
contribution to $\fcs$ is negligible at moderate values of $q$. Although
the precise numerical values of their contributions depend on one's
model for $\GMS$ and $\GES$, as well as on one's choice as to the
treatment of the retardation term, the overall magnitude of the
vector meson pair current contribution is sufficiently small so as to
render the impact of these model-dependencies negligible.

	In Figures 4 and 5, we plot the ratio $R_s=\fczs(q)/\fcztez(
q)$ $=\fcs(q)/\fctez(q)$,
which characterizes the $s$-quark corrections to the non-strange
PV asymmetry (Eqs. (3) and (6)). Assuming $|\GES/\GEn|\approx 1$
and $|\GMS/\GMTEZ|\approx 1$, which essentially
corresponds to assuming the
Jaffe values for $\rhostr$ and $\mustr$ but a
more realistic momentum-dependence in the strange form factors,
we expect a 35\% correction to the non-strange asymmetry (the first term
on the right side of Eq.~(6)) at the
kinematics of the CEBAF PV $^4$He experiment. Fig. 4 shows the dependence
of this correction on the value of $\mustr$ which, under our form factor
parameterization (Eqs. (12-13)), sets the scale of contributions
generated by $\GMS$. For purposes of illustration, we have assumed
magnitudes and relative signs
$\GES/\GEn\approx -1$ and $\GMS/\GETEZ\approx -1$ and have taken a positive
sign for $\gropiss$. The results at low-$q$ imply a negligible
dependence of $\fczs$ on $\mustr$ in this regime. For momentum
transfers in the vicinity of those suggested for the approved
CEBAF $^4$He experiment, the ratio changes by $\lapp$ 15\% as $\mustr$
is varied over a range of values suggested by model calculations [16-21].
Had we assumed the kaon loop values for $\rhostr$ and $\mustr$, the
overall size of the ratio $R_s$ would have been nearly ten times
smaller, so that in this case the relative impact of any uncertainty
in $\mustr$ would be correspondingly enhanced.

	Fig. 5 displays the impact on $R_s$ made by the choice of sign
of $\gropiss$, which one cannot determine from $\phi$-decay
data and symmetry arguments. We illustrate this sensitivity for two
different models of nucleon strangeness: (A) $(\rhostr, \mustr)=$
$(-2.0, -0.2)$ and (B) $(\rhostr, \mustr)=$ $(0.0, -0.2)$. For
low momentum-transfer, the impact of this uncertainty in sign is
negligible, whereas at $q\approx 4$ fm$^{-1}$ (the kinematics of
Ref.~[24]), it corresponds to roughly a $\pm 15\%$ uncertainty
in the asymmetry. To put the point somewhat differently, even if the
strange quark vector current matrix elements of the nucleon vanished
identically, we would expect non-nucleonic strange quarks in the nuclear
medium to generate a 15\% correction to the non-strange PV asymmetry
at the kinematics of Ref.~[24]. The scale of this effect is well below
the 40\% statistical error projected for the approved CEBAF experiment,
assuming 50\% beam polarization (the error is reduced to 28\% for 70\%
beam polarization). Thus, for a measurement of $\alr(^4{\rm He})$ to be
sensitive to $\gropis(Q^2)$, significantly a longer running time and/or
higher beam polarization would be required.

	In Fig. 6, we present the significance of a moderate-$q$
$\alr(^4{\rm He})$ measurement from a somewhat different perspective.
If one wishes to constrain the various strangeness parameters $\rhostr$,
$\mustr$, and $\lamsE$ at a level necessary to test model predictions in
detail, then a combination of experiments using proton and $A>1$ targets
would be required [1,2,22]. As noted in Refs.~[1,2,22], a combination of
low- and moderate-$q$ PV experiments with $^4$He could potentially constrain
$(\rhostr, \lamsE)$ more tightly than could a sequence of $\alr(\evec p)$
measurements alone. This conclusion was based on a one-body (IA)
calculation and the ideal assumption of 100\% beam polarization
with experimental errors being statistics dominated. The inclusion of
two-body currents does not alter our previous conclusion about the
possible constraints attainable from a low-$q$ measurement, since the
two-body contribution is negligible in this regime. In Fig. 6, we display
the impact of two-body currents on constraints attainable at moderate-$q$.
Fig. 6a shows the joint constraints on $(\rhostr, \mustr)$ a 10\%
measurement of $\alr(^4{\rm He})$ could produce, assuming
the parameterization of Eqs. (12-13) and central values for these
parameters given by model (A) discussed above. A similar plot for
$(\rhostr, \lamsE)$ constraints is given in Fig. 6b, where a central
value for $(\rhostr, \lamsE)=$ $(-2, \lamn)$ is assumed. The solid and
dashed lines give the constraints corresponding to different choices as
to the sign of $\gropiss$. We take the difference between these two sets
of lines as one measure of the theoretical uncertainty associated with our
calculation.

	From Fig. 6a, we note that the correlation between $\rhostr$
and $\mustr$ is weak. This feature follows from the relatively small
magnitudes of the vector meson pair current and spin-orbit contributions,
which carry the strongest dependences on $\mustr$. In the case of
Fig. 6b, we observe that the moderate-$q$ constraints are modified only
slightly from the IA expectation, even though many-body currents generate
significant contributions to $\fczs$ and $\fcztez$. The reason for the
insensitivity of these constraints to the two-body currents can be explained
in the following manner. First, the pionic corrections are proportional
to the Dirac form factor
$$
F_1^{(a)}=\left(1+\tau\right)^{-1}\left[\GEa+\tau\GMa\right]\ \ \ ,
\eqno(37)
$$
where \lq\lq $a$" denotes either the isoscalar EM current or
strange quark current. At the kinematics of the moderate-$q$ CEBAF
PV experiment, one has $\tau\approx 0.17$ so that $\tau\GMTEZ/\GETEZ
\approx 0.15$. In this case, $F_1^\sst{T=0}\approx\GETEZ$.
Similarly, $\tau\GMS/\GES\approx\mustr/\rhostr\approx 0.15$, assuming
the Jaffe values for the strangeness parameters, so that $\FOS\approx
\GES$. Under these assumptions, the pionic pair currents give the
dominant correction to the IA nuclear form factors, so that at
$\tau=0.17$ ($q^2=0.6$ $({\rm GeV}/c)^2$) one has
$$
\eqalignno{\fcztez(q)&\approx\rbra{{\rm g.s.}}\Mhat_0^\sst{T=0}(q)^{[1]}
	+\Mhat_0^\sst{T=0}(q)^{[2]}_{\rm pionic}\rket{{\rm g.s.}}&
	(38{\rm a})\cr
	&\approx\GETEZ(\tau)\rbra{{\rm g.s.}}{\hat{\cal O}}(q)^{[1]}+
	{\hat{\cal O}}(q)^{[2]}\rket{{\rm g.s.}}&\cr
	&\cr
	\fczs(q)&\approx\rbra{{\rm g.s.}}\Mhat_0^{(s)}(q)^{[1]}
	+\Mhat_0^{(s)}(q)^{[2]}_{\rm pionic}\rket{{\rm g.s.}}&
	(38{\rm b})\cr
	&\approx\GES(\tau)\rbra{{\rm g.s.}}{\hat{\cal O}}(q)^{[1]}+
	{\hat{\cal O}}(q)^{[2]}\rket{{\rm g.s.}}\ \ \ ,&\cr}
$$
where ${\hat{\cal O}}(q)^{[1]}$ and ${\hat{\cal O}}(q)^{[2]}$ are
nuclear operators (see, {\it e.g.}, Eqs. (15) and (A.1)). Hence, the ratio
$R_s=\fczs/\fcztez$ is essentially independent of nuclear matrix
elements and is governed by the ratio of single nucleon form
factors as in the IA case. Thus, the inclusion of two-body currents
does not seriously change the joint constraints on $\rhostr$ and
$\lamsE$. Some changes from the IA results do appear, since neither
$\GMS$ nor the $\rho$-$\pi$ constributions are completely negligible.
In the event that $|\mustr/\rhostr|>> 0.15$, however, this argument would
break down and our conclusions would have to be modified.

	We also point out that the uncertainty in the sign of the
$\rho$-$\pi$ transition current contribution does not seriously
affect the $(\rhostr, \lamsE)$ constraints, even though the magnitude
of this contribution is as large as the experimental uncertainty in $\alr$
assumed in obtaining the plots of Fig. 6.
To understand why this is the case, consider the following
argument. If one assumes that all of $\delta\alr$ translates into
an uncertainty $\delta R_s$ in the extracted value of $R_s$, and if
one further assumes that $\rhostr$ and $\lamsE$ can be varied in a manner
consistent with this uncertainty $\delta R_s$ (as we've done in
obtaining the lines in Fig. 6b), then one has
$$
{\delta\alr\over\alr}={\delta R_s\over 4\sstw+R_s}\ \ \ \eqno(98)
$$
or
$$
\delta R_s = \left(4\sstw+R_s\right){\delta\alr\over\alr}\approx
	0.1+0.1\ R_s\ \ \ \eqno(40)
$$
for $\delta\alr/\alr = 0.1$. Since $R_s$ changes by only $\pm 0.1$ for
different choices for the sign of $\gropiss$, the impact of this
choice on the magnitude of $\delta R_s$ -- and, therefore on the
joint constraints on $(\rhostr, \lamsE)$ -- is an order of magnitude
smaller than the impact of the experimental error in $\alr$.

	Finally, in Fig. 7 we show the $\rho$-meson pair current
contribution to $\fcs$ under the two different assumptions as to the
inclusion of the retardation correction. The curve labelled by
\lq\lq $\rho_\#$" was calculated without the
retardation correction (Gari-Hyuga convention [36]), while the curve
labelled by \lq\lq $\rho$" includes it (Riska
convention [32,33]).  The difference between the two should be taken
as an estimate of the theoretical uncertainty in the
treatment of these short-range currents.  Fortunately, the scale of
the vector meson contributions is
sufficiently small that the choice of convention has a negligible
impact on the value of $\fcs$.
\medskip
\noindent {\bf V. Conclusions.} We have computed MEC contributions to
the $^4$He strange quark elastic form factor,
$\fcs=2\sqrt{\pi}\fczs(q)$, using Monte
Carlo methods and an accurate variational ground state wavefunction. Our
results
indicate that the nuclear strangeness
radius, $\rhostr[{\rm nuc}]$, which governs $\fcs(q)$ at low
momentum-transfer, is (1) dominated by the single nucleon strangeness radius,
(2) two orders of magnitude less sensitive to many-body strangeness currents,
and (3) independent of pionic MEC's -- results which essentially confirm
our previous conclusions based on the shell model calculation.
At moderate values of $q$, such as
those corresponding to the approved CEBAF elastic PV $^4$He experiment
[24], we find that $\fcs$ generates a 35\% correction to the PV
asymmetry, assuming that $|\GES/\GEn|\approx 1$ and $|\GMS/\GMTEZ|\approx
1$ in this regime and that $\gropis(Q^2)$ is correctly given by
$\phi$-meson dominance. The magnitude of this correction is smaller than
the statistical error projected for the CEBAF experiment under the
most conservative assumptions about beam polarization. In the absence of
nucleonic strangeness, non-nucleonic $s\sbar$ pairs would generate
roughly a 15\% correction to the non-strange asymmetry at these
kinematics. Thus, the scale of the strange-quark contribution to
$\alr(^4{\rm He})$ is still sensitive to the nucleon's strangeness
electric form factor. In the event that $|\GES/\GEn|<<1$, a more
precise $^4$He PV measurement could probe the $\rho$-$\pi$ strangeness
charge operator. Such a measurement would be interesting since only
a $\rho$-$\pi$ transition three-current operator has been probed in
other experiments performed to date [27]. Finally, inclusion of
MEC contributions to $\fcs$ and $\fctez$ does not appear to affect
noticeably the constraints on the
leading and next-to-leading $Q^2$-dependence
of $\GES$ attainable with a medium-$q$ measurement of the $^4$He PV
asymmetry.
\medskip
\centerline{\bf Acknowledgements}
\medskip
We thank R.B. Wiringa for making available to us the latest version of
his variational $^4$He wavefunction.  We also thank J. Goity, E.J. Beise and
D.O. Riska
for useful discussions. This work is supported in part by funds provided
by the U.S. Department of Energy (D.O.E.) under contracts
\# DE-AC05-84ER40150 and \# DE-AC02-76ER03069. M. J. Musolf is supported
as a National Science Foundation Young Investigator.
\vfill
\eject
\centerline{\bf Appendix}
\medskip

In this Appendix, we provide complete expressions for the two-body
charge operators, expressions for the low-$q$ forms of the
corresponding Coulomb projections, and additional details of our
shell model calculation of the nuclear strangeness radius.
\medskip
\centerline{\bf Two-body charge operators}
\medskip
Expressions for the co-ordinate space charge operators can be
obtained by Fourier-transforming the momentum-space operators in
Eq. (22) and (24) and summing over all nucleon pairs.  The resulting
formulae are
$$
\eqalignno{\rohat(\qv)^{[2]}_{\rm pionic}&=
	-i\left[{\gpnns\over 16\pi m_\sst{N}^3}\right] F_1^{(a)}(\tau)
	\sum_{i<j}\delta(\xv_i-\xpv_i)\delta(\xv_j-\xpv_j)\tauv_i
	\cdot\tauv_j&(\hbox{A.1})\cr
	&\times\left[{\eexp{-\mpi\rij}\over r^2_{ij}}\right](1+\mpi\rij)
	\left[\eexp{i\qv\cdot\xv_i}\sigma_i \cdot \vec
q\,\sigv_j\cdot\rhij-\eexp{i\qv\cdot
	\xv_j}\sigv_j\cdot\qv\sigv_i\cdot\rhij\right]&\cr}
$$
for the pionic current,
$$
\eqalignno{\rohat(\qv)^{[2]\ \rm (a)}_{\rm V-pair}
	&=\left[{\gvnns\over 16 \pi m_\sst{N}^3}\right]
	G_\sst{M}^{(a)}(\tau)\sum_{i<j}\That_\sst{V}(i,j)
	\delta(\xv_i-\xpv_i)\delta(\xv_j-\xpv_j)&({\rm A.2a})\cr
	&\Biggl\{\left[{\eexp{-\mv\rij}\over\rij}\right]\left[\eexp{i
	\qv\cdot\xv_i}(\sigv_i\times\qv)\cdot\delvlr_j+\eexp{i\qv\cdot
	\xv_j}(\sigv_j\times\qv)\cdot\delvlr_i\right]&\cr
	&+i(1+\kapv)\left[{\eexp{-\mv\rij}\over r^2_{ij}}\right]
	(1+\mv\rij)\biggl[\eexp{i\qv\cdot\xv_j}(\qv\cdot\rhij+
	\sigv_j\times\qv\cdot\sigv_i\times\rhij)&\cr
	&\qquad -\eexp{i\qv\cdot\xv_i}(\qv\cdot\rhij+\sigv_i\times\qv
	\cdot\sigv_j\times\rhij)\biggr]\Biggr\}&\cr}
$$
for the vector meson pair current in the absence of the retardation
term and
$$
\eqalignno{\rohat(\qv)^{[2]\ \rm (b)}_{\rm V-pair}&=
	\rohat(\qv)^{[2]\ \rm (a)}_{\rm V-pair}
	+i\kapv(1+\kapv)\left[{\gvnns\over
	16 \pi m_\sst{N}^3}\right] F_1^{(a)}(\tau)&({\rm A.2b})\cr
	&\times\sum_{i<j}\That(i,j)
	\delta(\xv_i-\xpv_i)\delta(\xv_j-\xpv_j)
	\left[{\eexp{-\mv\rij}\over r^2_{ij}}\right]\left(
	1+\mv\rij\right)&\cr
	&\times\left[\eexp{i\qv\cdot\xv_j}\sigv_j\times
	\qv\cdot\sigv_i\times\rhij-\eexp{i\qv\cdot\xv_i}\sigv_i
	\times\qv\cdot\sigv_j\times\rhij\right]&\cr}
$$
with the retardation term included.
The $\rho$-$\pi$ transition current operator is given by
$$
\rohat(\qv)^{[2]}_{\rho\pi}=-\left[{\gpnn\gvnn\gropia(\tau)\over 32\pi\mns
	\mro}\right]\sum_{i<j}\delta(\xv_i-\xpv_i)\delta(\xv_j-\xpv_j)
	\tauv_i\cdot\tauv_j\eexp{i\qv\cdot\Rvij}\Gamma(i,j)\eqno(
	\hbox{A.3})
$$
where
$$
\eqalignno{\Gamma(i,j)&=i\left[F_1(\rvij)\delvlr_j\cdot(\rhij\times\qv)
	\sigv_i\cdot\qv-F_1(-\rvij)\delvlr_i\cdot(\rhij\times\qv)
	\sigv_j\cdot\qv\right]&(\hbox{A.4})\cr
	&+F_2(\rvij)\delvlr_j\cdot(\rhij\times\qv)\rhij\cdot\sigv_i
	+F_2(-\rvij)\delvlr_i\cdot(\rhij\times\qv)\rhij\cdot\sigv_j&\cr
	&+F_3(\rvij)\delvlr_j\cdot(\sigv_i\times\qv)+F_3(-\rvij)\delvlr_i
	\cdot(\sigv_j\times\qv)&\cr
	&+(1+\kapro)\biggl\{\sigv_i\cdot\sigv_j\left[i\left\{G_1(\rvij)
	+G_1(-\rvij)\right\}+\qv\cdot\rhij\left\{H_1(\rvij)-
	H_1(-\rvij)\right\}\right]&\cr
	&+\qv\cdot\sigv_i\qv\cdot\sigv_j\left[i\left\{G_2(\rvij)
	+G_2(-\rvij)\right\}+\qv\cdot\rhij\left\{H_2(\rvij)-
	H_2(-\rvij)\right\}\right]&\cr
	&+\rhij\cdot\sigv_i\rhij\cdot\sigv_j\left[i\left\{G_3(\rvij)
	+G_3(-\rvij)\right\}+\qv\cdot\rhij\left\{H_3(\rvij)-
	H_3(-\rvij)\right\}\right]&\cr
	&+\rhij\cdot\sigv_i\qv\cdot\sigv_j\left[\left\{G_4(\rvij)
	-G_5(-\rvij)\right\}+i\qv\cdot\rhij\left\{H_4(\rvij)+
	H_5(-\rvij)\right\}\right]&\cr
	&+\rhij\cdot\sigv_j\qv\cdot\sigv_i\left[\left\{G_5(\rvij)
	-G_4(-\rvij)\right\}+i\qv\cdot\rhij\left\{H_5(\rvij)+
	H_4(-\rvij)\right\}\right]\biggr\}&\cr}
$$
where
$$
\eqalignno{F_1(\rv)&=-\left(g_1+{g_0\over 2}\right)&(\hbox{A.5a})\cr
	   F_2(\rv)&=h_0+{g_0\over r}&\cr
	   F_3(\rv)&=-{g_0\over r}&\cr
	   &&\cr
	   G_1(\rv)&=-{1\over r}\left(g_1-{g_0\over 2}\right)q^2&(\hbox{A.5b})\cr
	   G_2(\rv)&={2g_1\over r}-h_1-{h_0\over 2}&\cr
	   G_3(\rv)&=q^2\left[{1\over r}\left(g_1-{g_0\over 2}\right)
		+\left(h_1-{h_0\over 2}\right)\right]&\cr
	   G_4(\rv)&=k_0-{h_0\over r}&\cr
	   G_5(\rv)&={h_0\over r}+q^2\left(g_2-{g_0\over 4}\right)&\cr
	   &&\cr
	   H_1(\rv)&={1\over r}\left(h_0+{g_0\over r}\right)&(\hbox{A.5c})\cr
	   H_2(\rv)&={g_0\over 4}-g_2&\cr
	   H_3(\rv)&=-\left(k_0+{3h_0\over r}+{3g_0\over r^2}\right)&\cr
	   H_4(\rv)&=-\left[\left(h_1-{h_0\over 2}\right)+{1\over r}
	   	\left(g_1-{g_0\over 2}\right)\right]&\cr
	   H_5(\rv)&={1\over r}\left(g_1+{g_0\over 2}\right)+\left(
	   	h_1+{h_0\over 2}\right)&\cr}
$$
and where
$$
\eqalignno{g_n(\rv,\qv)&=\int_{-1/2}^{1/2}\ d\beta \beta^n\exp\left[i\beta
	\qv\cdot\rv-Lr\right]&(\hbox{A.6})\cr
	&&\cr
	h_n(\rv, \qv)&=\int_{-1/2}^{1/2}\ d\beta \beta^n\ L\ \exp\left[i\beta
	\qv\cdot\rv-Lr\right]&\cr
	&&\cr
	k_n(\rv, \qv)&=\int_{-1/2}^{1/2}\ d\beta \beta^n\ L^2\ \exp\left[i\beta
	\qv\cdot\rv-Lr\right]&\cr}
$$
with
$$
L^2={1\over 2}(\mpis+\mros)+\beta(\mros-\mpis)+(1/4-\beta^2)q^2\ \ \ .
\eqno(\hbox{A.7})
$$
We define
$$
\delvlr_i={\overleftarrow\nabla}_i - {\overrightarrow\nabla}_i\ \ \ ,
\eqno(\hbox{A.8})
$$
where ${\overleftarrow\nabla}_i$ and ${\overrightarrow\nabla}_i$ are
gradients acting to the left and right, respectively, on the
co-ordinate of the $i$-th nucleon in the wavefunction (and not on
the co-ordinates appearing in the operators). The isospin tensor
$\That_\sst{V}(i,j)$ is defined in Eq.~(23), the quantities
$\xv_i$ and $\xpv_i$ are the co-ordinate of the $i$-th nucleon in
the initial and final state wavefunction, respectively, and where the
co-ordinates $\rvij$ {\it etc.} are defined in Eq.~(A.11) below. As
elsewhere, the superscript \lq\lq $a$" denotes either the $T=0$ EM
current or strangeness current operators.

		Expressions for the pair current operators with hadronic form
factors included (finite $\lamm$) may be obtained from the above formulae
by making the replacement
$$
\OP(\mm)\rightarrow\OP(\mm)-\OP(\lamm)+{(\lamms-\mms)\over 2\lamm}
{d\over d\lamm}\OP(\lamm)\ \ \ ,\eqno(\hbox{A.9})
$$
where $\OP(\mm)$ is any one of the pair current operators in Eqs.~(A.1,2)
associated with the exchange of a meson having mass $\mm$. Similarly,
for the $\rho$-$\pi$ transition current, the Coulomb operator in the presence
of hadronic form factors arises from making the replacement
$$
\OP(\mpi,\mro)\rightarrow\OP(\mpi,\mro)+\OP(\Lambda_\pi,\Lambda_\rho)
-\OP(\mpi,\Lambda_\rho)-\OP(\Lambda_\pi, \mro)\ \ \ ,\eqno(\hbox{A.10})
$$
where $\OP(\mpi,\mro)$ is the operator appearing in Eq.~(A.3).

	Substituting the above expressions for the charge operators into
Eq.~(7c) and expanding the exponentials in powers of $\qv$ leads to the
following expressions for the leading $q$-dependence of the Coulomb
operators:
$$
\eqalignno{\Mhat^{[2]}_{00}(q)\bigg\vert^{\rm pionic}_{q\to 0}&=\tau
\left[{\gpnns
\over 24\pi^{3/2}\mn}\right] F_1^{(a)}(\tau)\sum_{i<j}\delta(\xv_i-\xpv_i)
\delta(\xv_j-\xpv_j)\tauv_i\cdot\tauv_j&\cr
&&(\hbox{A.10a})\cr
&\times\left[{\eexp{-\mpi\rij}\over\mpi\rij}\right]\left(1+\mpi\rij\right)
\biggl[{1\over 3}\sigv_i\cdot\sigv_j&\cr
&\cr
&+\sqrt{8\pi\over3}[Y_2(\rhij)\otimes
[\sigma_i\otimes\sigma_j]_2]_0
+\left({\Rij\over\rij}\right)(\Rhij\times\rhij)\cdot(\sigv_i\times
\sigv_j)\biggr]\ \ \ ,&\cr}
$$
$$
\eqalignno{\Mhat^{[2]}_{00}(q)\bigg\vert^{V-{\rm pair}}_{q\to 0}&=\tau
\GMa(\tau)\left[{\gvnns\mv\over 24\pi^{3/2}\mn}\right]\sum_{i<j}\delta(
\xv_i-\xpv_i)\delta(\xv_j-\xpv_j)\That_\sst{V}(i,j)&\cr
&&(\hbox{A.10b})\cr
&\times\left[{\eexp{-\mv\rij}\over \mv\rij}\right]\Biggl\{
(1+\kapv)(1+\mv\rij)
\biggl[1+{2\over 3}\sigv_i\cdot\sigv_j&\cr
&\cr
&-\sqrt{8\pi\over 3}\left[Y_2(\rhij)
\otimes[\sigma_i\otimes\sigma_j]_2\right]_0&\cr
&\cr
&+\left({\Rij\over\rij}\right)(\rhij\times\Rhij)\cdot(\sigv_i\times
\sigv_j)\biggr]+{\vec\Sigma}_{ij}\cdot(\Lv_R^{ij}-2\Lv_r^{ij})&\cr
&\cr
&-i\vec\Delta_{ij}\cdot(\rhij\times\delv_\Rij-2\Rhij\times\delv_\rij)
-i\left({\Rij\over\rij}\right)(1+\mv\rij)\vec\Delta_{ij}\cdot
(\Rhij\times\rhij)\Biggr\}\ \ \ ,&\cr}
$$
$$
\eqalignno{\Mhat^{[2]}_{00}(q)\bigg\vert^{\rho\pi}_{q\to 0}&=\tau
\gropia(\tau)\left[{\gpnn\gronn\over 24\pi^{3/2}}\right]\sum_{i<j}\delta(
\xv_i-\xpv_i)\delta(\xv_j-\xpv_j)\tauv_i\cdot\tauv_j&\cr
&&(\hbox{A.10c})\cr
&\times{1\over\mro\rij}\left[{1\over (\mro\rij)^2-(\mpi\rij)^2}\right]
\Biggl\{(1+\kapro)\biggl[Z_1(i,j)\sigv_i\cdot\sigv_j&\cr
&\cr
&-\sqrt{8\pi\over 3}
Z_2(i,j)[Y_2(\rhij)\otimes[\sigma_i\otimes\sigma_j]_2]_0&\cr
&\cr
&+\left({\Rij\over\rij}\right)Z_3(i,j)(\rhij\times\Rhij)\cdot
(\sigv_i\times\sigv_j)\biggr]-{1\over 2}\rij Z_4(i,j)\delij
\cdot(\rhij\times\delv_\Rij)&\cr
&\cr
&+iZ_4(i,j)\sigij\cdot(\Lv_r^{ij}+\Lv_R^{ij})-2\Rij Z_4(i,j)
\delij\cdot(\Rhij\times\delv_\rij)&\cr
&\cr
&-2i\left({\Rij\over\rij}\right) Z_2(i,j)\rhij\cdot\delij\Rhij\cdot
\Lv_r^{ij}-iZ_2(i,j)\sigij\cdot\rhij \rhij\cdot\Lv_R^{ij}\Biggr\}
\ \ \ ,&\cr}
$$
where
$$
\eqalignno{\rvij&=\xv_i-\xv_j&\cr
\rij&=|\rvij|&(\hbox{A.11a})\cr
\rhij&=\rvij/\rij&\cr
&\cr
\Rvij&={1\over 2}(\xv_i+\xv_j)&(\hbox{A.11b})\cr
\Rij&=|\Rvij|&\cr
&\cr
\Lv_r^{ij}&=-i\rvij\times\delv_\rij&(\hbox{A.11c})\cr
\Lv_R^{ij}&=-i\Rvij\times\delv_\Rij&\cr
&\cr
\sigij&=\sigv_i+\sigv_j&(\hbox{A.11d})\cr
\delij&=\sigv_i-\sigv_j\ \ \ ,&\cr}
$$
and where
$$
\eqalignno{Z_1(i,j)&=(\mpi\rij)^2\eexp{-\mpi\rij}-(\mro\rij)^2
\eexp{-\mro\rij}&(\hbox{A.12a})\cr
&\cr
Z_2(i,j)&=\left[3+3(\mro\rij)+(\mro\rij)^2\right]
\eexp{-\mro\rij}&(\hbox{A.12b})\cr
&-\left[3+3(\mpi\rij)+(\mpi\rij)^2\right]\eexp{-\mpi\rij}&\cr
&\cr
Z_3(i,j)&=\left[2+2(\mro\rij)+
(\mro\rij)^2+(\mro\rij)^3\right]\eexp{-\mro\rij}
&\cr
&-\left[2+2(\mpi\rij)+(\mpi\rij)^2+(\mpi\rij)^3\right]\eexp{-\mro\rij}&(
\hbox{A.12c})\cr
&\cr
Z_4(i,j)&=\left[1+\mro\rij\right]\eexp{-\mro\rij}-\left[1+\mpi\rij\right]
\eexp{-\mpi\rij}\ \ \ .&(\hbox{A.12d})\cr}
$$
We note in passing that the overall normalization of the $\rho-\pi$ operator
appearing in Eq.~(A.10c) differs by a factor of four from that appearing in
Eq.~(7) of Ref.~[26]. The latter, as well as the terms involving
$\gropis$ in Eqs.~(10) and (14) of that work, should be multiplied by
$1/4$.
\medskip
\centerline{\bf Shell Model Calculation}
\medskip

	Use of a simple shell model $^4$He ground state allows one to
obtain analytic expressions for the nuclear strangeneness radius,
$\rhostr[{\rm nuc}]$, which are useful in the interpretation of the
numerical results obtained with variational ground state wavefunctions.
To that end, we compute matrix elements of the one- and two-body Coulomb
operators in the low-$q$ limit. From the expressions in Eqs.~(A.10) and
as noted in the main text of the paper, the vector meson pair current
and $\rho$-$\pi$ transition current Coulomb operators go as $q^2$ for
low-$q$. The two-body pionic operator, in contrast, vanishes as
$q^4$ since $F_1^{(s)}\sim q^2$ and since the operator carries an
additional, explicit factor of $q^2$. Similarly, the vector meson
pair current retardation term also enters at ${\cal O}(q^4)$. Thus,
for purposes of computing two-body contributions to $\rhostr[{\rm nuc}]$,
we need only compute matrix elements of the operators in Eqs.~(A.10b,c).
In the limit that the $^4$He ground state is a pure S-state, the
leading $q^2$-dependence of the one-body matrix element (Eq.~(15)) is given
by the one-body form factor times the number of nucleons and is
independent of details of the nuclear wavefunction. The two-body matrix
elements, on the other hand, are structure-dependent.
An important consideration in this respect is the
short-range repulsion between nucleons. Since the ranges of the $\rho$-
and $\omega$-mesons are commensurate with the radius of the repulsive
core in the N-N potential, matrix elements of the vector meson exchange
operators ought to be suppressed. To account for this effect, we compute
the two-body shell model matrix elements by including a phenomenological
correlation function, $g(r)$, in the integral over relative co-ordinates:
$$
\int_0^\infty\ r^2dr\ u^*(r)\OP u(r)\rightarrow \int_0^\infty\ r^2dr\
g(r)\ u^*(r)\OP u(r)\ \ \ ,\eqno(\hbox{A.13})
$$
where $u(r)$ is the radial wavefunction for the relative motion of two
nucleons, $r=|\xv_1-\xv_2|$ is the relative co-ordinate, and $\OP$ is
an $r$-dependent two-body operator. Following the approach of Ref.~[34],
we take the correlation function to have the form
$$
g(r)=C\left[1-\eexp{-r^2/d^2}\right]\ \ \ ,\eqno(\hbox{A.14})
$$
where the constant $C$ is determined by the requirement that the wavefunction
be normalized. A fit to the nuclear matter
correlation function of Ref.~[47]
gives $d=0.84$ fm. With this form for $g(r)$, the un-correlated two-body
matrix elements are modified as
$$
\hcal{M}(b)\rightarrow C\left[\hcal{M}(b)-\hcal{M}(\beff)\right]\ \ \ ,
\eqno(\hbox{A.15})
$$
where $\hcal{M}(b)$ is the un-correlated two-body matrix element computed
using an oscillator parameter $b$, where the effective oscillator parameter
is given by
$$
\left({\beff\over b}\right) = \left[1+2\left({b\over d}\right)\right]^{-1/2}
\ \ \ ,\eqno(\hbox{A.16})
$$
and where
$$
C=\left[1-\left({\beff\over b}\right)^3\right]^{-1}\ \ \ .\eqno(\hbox{A.17})
$$

In the limit of no short-range repulsion ($\beff\to 0$), one has for
the leading $q$-dependence of $\fczs$ the expression given in Eq.~(26).
The nuclear $\lambda_{1,2}$ are
given by
$$
\eqalignno{\lambda_1&=2/\sqrt{\pi}\ \ \ ,&(\hbox{A.18a})\cr
&\cr
\lambda_{2a}&=-\sum_{V=\rho, \omega}(1+\kapv)\left({\sqrt{2}
\gvnns\over 24\pi^2}\right)\left({\mv\over\mn}\right)
{{\cal N}_\sst{V}\over (\mv b)}&(\hbox{A.18b})\cr
&\times\left\{1-(\mv b)^2+\sqrt{\pi/2}(\mv b)^3\exp\left[\coeff{1}{2}(\mv b)^2
\right]{\rm erfc}\left({\mv b\over\sqrt{2}}\right)\right\}&\cr
&\rightarrow -\sum_{V=\rho, \omega}(1+\kapv)\left({\sqrt{2}
\gvnns\over 8\pi^2}\right)\left({\mv\over\mn}\right)
{{\cal N}_\sst{V}\over (\mv b)^3}
\left[1-{5\over (\mv b)^2}+\cdots\right]\ \ \ ,&\cr
&\cr
\lambda_{2b}&=(1+\kapro)\left({\sqrt{2}\gpnn\gronn\over 18\pi^2}\right)
{{\cal N}_2\over (\mro b)}\left[{1\over (\mro b)^2-(\mpi
b)^2}\right]&(\hbox{A.18c})\cr
&\cr
&\times\left[(\mpi b)^2 I(\mpi b) - (\mro b)^2 I(\mro b)\right]\ \ \ ,&\cr}
$$
where
$$
I(mb)=1-\sqrt{\pi\over 2} (mb)\exp[\coeff{1}{2}(mb)^2]{\rm erfc}\left({mb
\over\sqrt{2}}\right)\ \ \ ,\eqno(\hbox{A.19})
$$
and where ${\cal N}_{\sst{V},2}$ are spin-isospin matrix elements and
$\gvnn$ is the vector meson-nucleon coupling. For $b=1.2$ fm, one has
$\mro b >> 1$, so that the function in Eq.~(A.18c) may be expanded as
$$
(\mro b)^2 I(\mro b)= 1-{3\over (\mro b)^2}+\cdots\ \ \ .\eqno(\hbox{A.20})
$$
A similar expansion in powers of $1/(\mv b)$ has been used in arriving
at the expression in Eq.~(A.18b), where the $+\cdots$ indicates contributions
from terms higher order in $1/(\mv b)$.

\vfill
\eject
\medskip
\centerline{\bf References}
\medskip
\item{1.}M. J. Musolf {\it et al.}, {\it Physics Reports\bf\ 239/1\&2}
(1994).
\medskip
\item{2.}M. J. Musolf and T.W. Donnelly, \NPA{546} (1992) 509; \NPA{550}
	(1992) 564 (E).
\medskip
\item{3.}W. M. Alberico {\it et al.}, \NPA{569} (1994) 701.
\medskip
\item{4.}T. W. Donnelly {\it et al.}, \NPA{541} (1992) 525.
\medskip
\item{5.}D. H. Beck, \PRD{39} (1989) 3248.
\medskip
\item{6.}MIT-Bates proposal \# 89-06, R. D. McKeown and D. H. Beck,
contact people.
\medskip
\item{7.}D. B. Kaplan and A. Manohar, \NPB{310} (1988) 527.
\medskip
\item{8.}T. P. Cheng, \PRD{13} (1976) 2161.
\medskip
\item{9.}J. Gasser, H. Leutwyler, and M. E. Sainio, \PLB{253} (1991) 252.
\medskip
\item{10.}L.A. Ahrens {\it et al.}, {\it Phys. Rev. \bf D35} (1987) 785.
\medskip
\item{11.}G. Garvey {\it et al.}, \PRC{48} (1993) 1919.
\medskip
\item{12.}C. J. Horowitz {\it et al.}, \PRC{48} (1993) 3078.
\medskip
\item{13.}J. Ashman {\it et al.}, {\it Nucl. Phys. \bf B328} (1989) 1.
\medskip
\item{14.}P. L. Anthony {\it et al.}, \PLB{302} (1993) 553.
\medskip
\item{15.}B. Adeva {\it et al.}, \PRL{71} (1993) 959.
\medskip
\item{16.}R. L. Jaffe, {\it Phys. Lett. \bf B229} (1989) 275.
\medskip
\item{17.}N. W. Park, J. Schechter and H. Weigel, {\it Phys. Rev.\/}
{\bf D43}, 869 (1991).
\medskip
\item{18.} B. R. Holstein in {\it Proceedings of the Caltech Workshop on
	Parity Violation in Electron Scattering}, E.J. Beise and R.D.
	McKeown, Eds., World Scientific (1990) pp. 27-43.
\medskip
\item{19.} W. Koepf, E.M. Henley, and S.J. Pollock, \PLB{288} (1992) 11.
\medskip
\item{20.}M. J. Musolf and M. Burkardt, \ZPC{61} (1994) 433.
\medskip
\item{21.}T. Cohen, H. Forkel, and M. Nielsen, \PLB{316} (1993) 1.
\medskip
\item{22.}M. J. Musolf and T. W. Donnelly, \ZPC{57} (1993) 559.
\medskip
\item{23.}J.D. Walecka in {\sl Muon Physics}, Vol. II, V.W. Hughes and
	C.S. Wu, Eds., Academic Press (1975) 113.
\medskip
\item{24.}CEBAF proposal \# PR-91-004, E.J. Beise, spokesperson.
\medskip
\item{25.}CEBAF proposal \# PR-91-017, D.H. Beck, spokesperson.
\medskip
\item{26.}M. J. Musolf and T. W. Donnelly, \PLB{318} (1993) 263.
\medskip
\item{27.}E. Hummel and J.A. Tjon, {\it Phys. Rev. Lett} {\bf 63} (1989) 1788.
\medskip
\item{28.}R.G. Sachs, {\it Phys. Rev. \bf 126} (1962) 2256.
\medskip
\item{29.}T. de~Forest, Jr. and J. D. Walecka, {\it Adv. in Phys.\/} {\bf 15},
(1966) 1.
\medskip
\item{30.}J. D. Bjorken and S. D. Drell, {\sl Relativistic Quantum
Mechanics}, McGraw-Hill (New York, 1964) Chapter 3.
\medskip
\item{31.}CEBAF proposal \# PR-91-017, D.H. Beck, spokesperson.
\medskip
\item{32.}R. Schiavilla, V. R. Pandharipande, and D. O. Riska,
	\PRC{41} (1990)309.
\medskip
\item{33.}D. O. Riska, {\it Prog. Part. Nucl. Phys.\bf 11} (1984) 199.
\medskip
\item{34.}J. Dubach, J. H. Koch, and T. W. Donnelly, \NPA{271} (1976) 279.
\medskip
\item{35.}M. Chemtob and M. Rho, \NPA{163} (1971) 1.
\medskip
\item{36.}M. Gari and H. Hyuga, \NPA{264} (1976) 409.
\medskip
\item{37.}G. H\"ohler and E. Pietarinen, {\it Nucl. Phys.} {\bf B95} (1975)
210.
\medskip
\item{38.}The overal minus sign on the right hand side of Eq. (18) was
inadvertantly omitted from Eq.~(6) of Ref. [26].
\medskip
\item{39.}D. Berg {\it et al.}, {\it Phys. Rev. Lett.} {\bf 44} (1980) 706.
\medskip
\item{40.}J. L. Goity and M. J. Musolf, to be published.
\medskip
\item{41.} R. B. Wiringa, R. A. Smith and T. L. Ainsworth, \PRC{29} (1984)
1207.
\medskip
\item{42.} R. B. Wiringa, \PRC{43} (1991) 1585.
\medskip
\item{43.} R. B. Wiringa, private communication.
\medskip
\item{44.} R. Schiavilla, V. R. Pandharipande and D. O. Riska, \PRC{40} (1989)
2294.
\medskip
\item{45.} R. Schiavilla and D. O. Riska, \PRC{43} (1991) 437.
\medskip
\item{46.} R. Machleidt, Adv. Nucl. Phys. {\bf 19} (1989) 1.
\medskip
\item{47.} E. J. Moniz and G. D. Nixon, \AoP{67} (1971) 58.
\vfill
\eject
\centerline{\bf Figure Captions}
\bigskip
\noindent{\bf Fig. 1.} Two-nucleon ($N$ and $N'$)
meson exchange current (MEC) contributions
to nuclear matrix elements of the isoscalar EM and strange-quark
vector currents. \lq\lq Pair current" processes, shown in (a,b),
arise from the negative energy pole in the intermediate-state, single
nucleon propagator. \lq\lq Transition current" contributions (c)
are generated by matrix element of the current operators (indicated
by the $\otimes$) between mesonic states ($M'$ and $M$). As explained
in the text, mesonic matrix elements of $J_\mu^\sst{EM}(T=0)$ and $\sbar
\gamma_\mu s$ vanish when $M'=M$.
\medskip
\noindent{\bf Fig. 2.} $^4$He elastic charge form factor,
$\fctez(q)$. Panel (a) gives the absolute value of the
form factor.
Circles indicate experimental values. Dashed curve
gives theoretical prediction in the impulse approximation (IA)
while the solid curve results from the inclusion of two-body
currents (IA+MEC). Panel (b) shows individual
one- and two-body contributions to $\fctez(q)$.
One-body contribution is indicatd by solid curve (IA).
Dashed curves give contributions from
the pionic (circles), $\rho$-meson (squares),
and $\omega$-meson (asterisks) pair currents as well as the
$\rho$-$\pi$ \lq\lq transition current (triangles).  Short dashed
curve indicates the spin-orbit contribution. Only the absolute value
of each contribution is plotted, and the signs are indicated in
parenthesis. Vector meson pair current contributions are computed
including the retardation correction to the charge operator (Eqs.
22c and A.2b).
\medskip
\noindent{\bf Fig. 3.} Same as Fig. 2, but for the elastic
strangeness charge form factor of $^4$He, $\fcs(q)$. In this case,
only theoretical predictions are shown, since no measurements have
as yet been made. Computations were carried out using $(\rhostr, \mustr)$
$=$ $(-2.0, -0.2)$, which correspond roughly to the pole model
predictions for these parameters [16], and a Galster-like parameterization
for the $q$-dependence of the one-body strangeness form factors.
A positive sign for $\gropiss$ was also assumed.
\medskip
\noindent{\bf Fig. 4.} Elastic strangeness to EM charge form factor
ratio, $\fcs(q)/\fctez(q)$ for different values of nucleon strangeness
magnetic moment, $\mustr$, and fixed strangeness radius, $\rhostr$.
\medskip
\noindent{\bf Fig. 5.} Same as Fig. 4, but for fixed $\mustr$ and
variable $\rhostr$. In each case, results using the impulse approximation
for $\fcs$ are shown along with results including two-body currents
for two different choices of sign on $\gropiss$. Panel (a) assumes
a large negative value for $\rhostr$, while panel (b) gives the ratio
for vanishing nucleon strangeness radius.
\medskip
\noindent{\bf Fig. 6.} Prospective constraints on single nucleon
strangeness parameters from a 10\% measurement of the $^4$He elastic,
PV asymmetry at $Q^2=-0.6$ $(\hbox{GeV}/c)^2$. Solid (dashed) lines
give the band of allowed values for positive (negative) sign on
$\gropiss$. In panel (a), $\lamsE$ is assumed fixed, while in panel (b),
$\mustr$ is held constant. In both panels, central values of $(\rhostr,
\mustr, \lamsE)$ $=$ $(-2.0, -0.2, 5.6)$ are assumed for purposes of
illustration.
\medskip
\noindent{\bf Fig. 7.} $\rho$-meson pair current contribution to $\fcs(q)$
computed including the retardation correction and omitting it (\lq\lq \#"
subscript). Only absolute value is plotted, while sign is indicated in
parenthesis.
\vfill
\eject
\end
%
%
%
%
%

\def\gboxit#1{\hbox{\vrule\vbox{\hrule\kern3pt\vtop
{\hbox{\kern3pt#1\kern3pt}
\kern3pt\hrule}}\vrule}}

\def\ttilde#1{\raise2ex\hbox{${\scriptscriptstyle(}\!
\sim\scriptscriptstyle{)}$}\mkern-16.5mu #1}
\def\dddots#1{\raise1ex\hbox{$^{\ldots}$}\mkern-16.5mu #1}
\def\pp#1#2{\raise1.5ex\hbox{${#2}$}\mkern-17mu #1}
\def\upleftarrow#1{\raise2ex\hbox{$\leftarrow$}\mkern-16.5mu #1}
\def\uprightarrow#1{\raise2ex\hbox{$\rightarrow$}\mkern-16.5mu #1}
\def\upleftrightarrow#1{\raise1.5ex\hbox{$\leftrightarrow$}\mkern-16.5mu #1}
\def\bx#1#2{\vcenter{\hrule \hbox{\vrule height #2in \kern #1in\vrule}\hrule}}

\def\squiggle#1{\lower1.5ex\hbox{$\sim$}\mkern-14mu #1}

\def\narrower{\advance\leftskip by\parindent \advance\rightskip by\parindent}


\def\mbox#1#2{\vcenter{\hrule width#1in\hbox{\vrule height#2in
   \hskip#1in\vrule height#2in}\hrule width#1in}}
\def\eqsquare #1:#2:{\vcenter{\hrule width#1\hbox{\vrule height#2
   \hskip#1\vrule height#2}\hrule width#1}}
\def\inbox#1#2#3{\vcenter to #2in{\vfil\hbox to #1in{$$\hfil#3\hfil$$}\vfil}}
\def\strutdepth{\dp\strutbox}
\def\marbul{\strut\vadjust{\kern-\strutdepth\specialbul}}
\def\specialbul{\vtop to \strutdepth{
    \baselineskip\strutdepth\vss\llap{$\bullet$\qquad}\null}}
\def\Bcomma{\lower6pt\hbox{$,$}}    
\def\bcomma{\lower3pt\hbox{$,$}}    

\def\sl{\scrsf}

\def\updots{\mathinner{\mskip 1mu\raise 1pt\hbox{.}
    \mskip 2mu\raise 4pt\hbox{.}\mskip 2mu
    \raise 7pt\vbox{\kern 7pt\hbox{.}}\mskip 1mu}}

\def\square{\kern1pt\vbox{\hrule height 1.2pt\hbox{\vrule width 1.2pt\hskip 3pt
   \vbox{\vskip 6pt}\hskip 3pt\vrule width 0.6pt}\hrule height 0.6pt}\kern1pt}
\def\ssquare{\kern1pt\vbox{\hrule height .6pt\hbox{\vrule width .6pt\hskip 3pt
   \vbox{\vskip 6pt}\hskip 3pt\vrule width 0.6pt}\hrule height 0.6pt}\kern1pt}
\def\lege{\hbox{$ {     \lower.40ex\hbox{$>$}
                   \atop \raise.20ex\hbox{$<$}
                   }     $}  }

\def\rege{\hbox{$ {     \lower.40ex\hbox{$<$}
                   \atop \raise.20ex\hbox{$>$}
                   }     $}  }

\def\lapp{\hbox{$ {     \lower.40ex\hbox{$<$}
                   \atop \raise.20ex\hbox{$\sim$}
                   }     $}  }
\def\rapp{\hbox{$ {     \lower.40ex\hbox{$>$}
                   \atop \raise.20ex\hbox{$\sim$}
                   }     $}  }

\def\tridots{\hbox{$ {     \lower.40ex\hbox{$.$}
                   \atop \raise.20ex\hbox{$.\,.$}
                   }     $}  }
\def\Times{\times\hskip-2.3pt{\raise.25ex\hbox{{$\scriptscriptstyle|$}}}}

\def\rightonleft{\hbox{$ {     \lower.40ex\hbox{$\longrightarrow$}
                   \atop \raise.20ex\hbox{$\longleftarrow$}
                   }     $}  }

\def\pmb#1{\setbox0=\hbox{$#1$}%
\kern-.025em\copy0\kern-\wd0
\kern.05em\copy0\kern-\wd0
\kern-.025em\raise.0433em\box0 }

%
%
\font\fivebf=cmbx5
\font\sixbf=cmbx6
\font\sevenbf=cmbx7
\font\eightbf=cmbx8
\font\ninebf=cmbx9
\font\tenbf=cmbx10

\font\bfmone=cmbx10 scaled\magstep1

\font\sevenit=cmti7
\font\eightit=cmti8
\font\nineit=cmti9
\font\tenit=cmti10

\font\itmone=cmti10 scaled\magstep1

\font\fiverm=cmr5
\font\sixrm=cmr6
\font\sevenrm=cmr7
\font\eightrm=cmr8
\font\ninerm=cmr9
\font\tenrm=cmr10

\font\rmmone=cmr10 scaled\magstep1

\def\fontone{\def\rm{\fcm0\rmmone}%
  \textfont0=\rmmone \scriptfont0=\tenrm \scriptscriptfont0=\sevenrm
  \textfont1=\itmone \scriptfont1=\tenit \scriptscriptfont1=\sevenit
  \def\it{\fcm\itfcm\itmone}%
  \textfont\itfcm=\itmone
  \def\bf{\fcm\bffcm\bfmone}%
  \textfont\bffcm=\bfmone \scriptfont\bffcm=\tenbf
   \scriptscriptfont\bffcm=\sevenbf
  \tt \ttglue=.5em plus.25em minus.15em
  \normalbaselineskip=25pt
  \let\sc=\tenrm
  \let\big=\tenbig
  \setbox\strutbox=\hbox{\vrule height10.2pt depth4.2pt width\z@}%
  \normalbaselines\rm}



\font\ninerm=cmr9
\font\eightrm=cmr8
\font\sixrm=cmr6

\font\ninei=cmmi9
\font\eighti=cmmi8
\font\sixi=cmmi6
\skewchar\ninei='177 \skewchar\eighti='177 \skewchar\sixi='177

\font\ninesy=cmsy9
\font\eightsy=cmsy8
\font\sixsy=cmsy6
\skewchar\ninesy='60 \skewchar\eightsy='60 \skewchar\sixsy='60

\font\ninebf=cmbx9
\font\eightbf=cmbx8
\font\sixbf=cmbx6

\font\ninett=cmtt9
\font\eighttt=cmtt8

\hyphenchar\tentt=-1 
\hyphenchar\ninett=-1
\hyphenchar\eighttt=-1

\font\ninesl=cmsl9
\font\eightsl=cmsl8

\font\nineit=cmti9
\font\eightit=cmti8


\newskip\ttglue
\def\tenpoint{\def\rm{\fcm0\tenrm}%
  \textfont0=\tenrm \scriptfont0=\sevenrm \scriptscriptfont0=\fiverm
  \textfont1=\teni \scriptfont1=\seveni \scriptscriptfont1=\fivei
  \textfont2=\tensy \scriptfont2=\sevensy \scriptscriptfont2=\fivesy
  \textfont3=\tenex \scriptfont3=\tenex \scriptscriptfont3=\tenex
  \def\it{\fcm\itfcm\tenit}%
  \textfont\itfcm=\tenit
  \def\sl{\fcm\slfcm\tensl}%
  \textfont\slfcm=\tensl
  \def\bf{\fcm\bffcm\tenbf}%
  \textfont\bffcm=\tenbf \scriptfont\bffcm=\sevenbf
   \scriptscriptfont\bffcm=\fivebf
  \def\tt{\fcm\ttfcm\tentt}%
  \textfont\ttfcm=\tentt
  \tt \ttglue=.5em plus.25em minus.15em
  \normalbaselineskip=16pt
  \let\sc=\eightrm
  \let\big=\tenbig
  \setbox\strutbox=\hbox{\vrule height8.5pt depth3.5pt width\z@}%
  \normalbaselines\rm}

\def\ninepoint{\def\rm{\fcm0\ninerm}%
  \textfont0=\ninerm \scriptfont0=\sixrm \scriptscriptfont0=\fiverm
  \textfont1=\ninei \scriptfont1=\sixi \scriptscriptfont1=\fivei
  \textfont2=\ninesy \scriptfont2=\sixsy \scriptscriptfont2=\fivesy
  \textfont3=\tenex \scriptfont3=\tenex \scriptscriptfont3=\tenex
  \def\it{\fcm\itfcm\nineit}%
  \textfont\itfcm=\nineit
  \def\sl{\fcm\slfcm\ninesl}%
  \textfont\slfcm=\ninesl
  \def\bf{\fcm\bffcm\ninebf}%
  \textfont\bffcm=\ninebf \scriptfont\bffcm=\sixbf
   \scriptscriptfont\bffcm=\fivebf
  \def\tt{\fcm\ttfcm\ninett}%
  \textfont\ttfcm=\ninett
  \tt \ttglue=.5em plus.25em minus.15em
  \normalbaselineskip=11pt
  \let\sc=\sevenrm
  \let\big=\ninebig
  \setbox\strutbox=\hbox{\vrule height8pt depth3pt width\z@}%
  \normalbaselines\rm}

\def\eightpoint{\def\rm{\fcm0\eightrm}%
  \textfont0=\eightrm \scriptfont0=\sixrm \scriptscriptfont0=\fiverm
  \textfont1=\eighti \scriptfont1=\sixi \scriptscriptfont1=\fivei
  \textfont2=\eightsy \scriptfont2=\sixsy \scriptscriptfont2=\fivesy
  \textfont3=\tenex \scriptfont3=\tenex \scriptscriptfont3=\tenex
  \def\it{\fcm\itfcm\eightit}%
  \textfont\itfcm=\eightit
  \def\sl{\fcm\slfcm\eightsl}%
  \textfont\slfcm=\eightsl
  \def\bf{\fcm\bffcm\eightbf}%
  \textfont\bffcm=\eightbf \scriptfont\bffcm=\sixbf
   \scriptscriptfont\bffcm=\fivebf
  \def\tt{\fcm\ttfcm\eighttt}%
  \textfont\ttfcm=\eighttt
  \tt \ttglue=.5em plus.25em minus.15em
  \normalbaselineskip=9pt
  \let\sc=\sixrm
  \let\big=\eightbig
  \setbox\strutbox=\hbox{\vrule height7pt depth2pt width\z@}%
  \normalbaselines\rm}

\font\scrsf=cmssi10                

\font\ninebf=cmbx9




\def\gboxit#1{\hbox{\vrule\vbox{\hrule\kern3pt\vtop
{\hbox{\kern3pt#1\kern3pt}
\kern3pt\hrule}}\vrule}}

\def\ttilde#1{\raise2ex\hbox{${\scriptscriptstyle(}\!
\sim\scriptscriptstyle{)}$}\mkern-16.5mu #1}
\def\dddots#1{\raise1ex\hbox{$^{\ldots}$}\mkern-16.5mu #1}
\def\siton#1#2{\raise1.5ex\hbox{$\scriptscriptstyle{#2}$}\mkern-17.5mu #1}
\def\pp#1#2{\raise1.5ex\hbox{${#2}$}\mkern-17mu #1}
\def\upleftarrow#1{\raise2ex\hbox{$\leftarrow$}\mkern-16.5mu #1}
\def\uprightarrow#1{\raise2ex\hbox{$\rightarrow$}\mkern-16.5mu #1}
\def\upleftrightarrow#1{\raise1.5ex\hbox{$\leftrightarrow$}\mkern-16.5mu #1}
\def\bx#1#2{\vcenter{\hrule \hbox{\vrule height #2in \kern #1in\vrule}\hrule}}

\def\squiggle#1{\lower1.5ex\hbox{$\sim$}\mkern-14mu #1}

\def\narrower{\advance\leftskip by\parindent \advance\rightskip by\parindent}

\def\onsim{\hbox{$ {     \lower.40ex\hbox{$\sim$}
                   \atop \raise.20ex\hbox{$+$}
                   }     $}  }

\def\simon{\hbox{$ {     \lower.40ex\hbox{$+$}
                   \atop \raise.20ex\hbox{$\sim$}
                   }     $}  }


\def\mbox#1#2{\vcenter{\hrule width#1in\hbox{\vrule height#2in
   \hskip#1in\vrule height#2in}\hrule width#1in}}
\def\eqsquare #1:#2:{\vcenter{\hrule width#1\hbox{\vrule height#2
   \hskip#1\vrule height#2}\hrule width#1}}
\def\inbox#1#2#3{\vcenter to #2in{\vfil\hbox to #1in{$$\hfil#3\hfil$$}\vfil}}
\def\strutdepth{\dp\strutbox}
\def\marbul{\strut\vadjust{\kern-\strutdepth\specialbul}}
\def\specialbul{\vtop to \strutdepth{
    \baselineskip\strutdepth\vss\llap{$\bullet$\qquad}\null}}
\def\Bcomma{\lower6pt\hbox{$,$}}    
\def\bcomma{\lower3pt\hbox{$,$}}    

\def\sl{\scrsf}

\def\updots{\mathinner{\mskip 1mu\raise 1pt\hbox{.}
    \mskip 2mu\raise 4pt\hbox{.}\mskip 2mu
    \raise 7pt\vbox{\kern 7pt\hbox{.}}\mskip 1mu}}

\def\square{\kern1pt\vbox{\hrule height 1.2pt\hbox{\vrule width 1.2pt\hskip 3pt
   \vbox{\vskip 6pt}\hskip 3pt\vrule width 0.6pt}\hrule height 0.6pt}\kern1pt}
\def\ssquare{\kern1pt\vbox{\hrule height .6pt\hbox{\vrule width .6pt\hskip 3pt
   \vbox{\vskip 6pt}\hskip 3pt\vrule width 0.6pt}\hrule height 0.6pt}\kern1pt}
\def\lege{\hbox{$ {     \lower.40ex\hbox{$>$}
                   \atop \raise.20ex\hbox{$<$}
                   }     $}  }

\def\rege{\hbox{$ {     \lower.40ex\hbox{$<$}
                   \atop \raise.20ex\hbox{$>$}
                   }     $}  }

\def\tridots{\hbox{$ {     \lower.40ex\hbox{$.$}
                   \atop \raise.20ex\hbox{$.\,.$}
                   }     $}  }
\def\Times{\times\hskip-2.3pt{\raise.25ex\hbox{{$\scriptscriptstyle|$}}}}

\def\rightonleft{\hbox{$ {     \lower.40ex\hbox{$\longrightarrow$}
                   \atop \raise.20ex\hbox{$\longleftarrow$}
                   }     $}  }

\def\pmb#1{\setbox0=\hbox{$#1$}%
\kern-.025em\copy0\kern-\wd0
\kern.05em\copy0\kern-\wd0
\kern-.025em\raise.0433em\box0 }